\documentclass{jfm}
\usepackage{graphicx,color}
\usepackage{float}
\usepackage{epstopdf, epsfig}
\usepackage{caption,subcaption}
\usepackage[intlimits]{amsmath}
\usepackage{lscape}
\usepackage{rotating}
\usepackage{amssymb, amsmath}
\usepackage{amsmath}
\usepackage[mathscr]{euscript}
\usepackage{soul}
\usepackage{hyperref} % For the \url command
\newcommand{\CRCA}[1]{{\textcolor{black}{#1}}}
\newcommand{\CRCAr}[1]{{\textcolor{black}{#1}}}

\usepackage[normalem]{ulem}  % In your preamble

\DeclareRobustCommand{\varlambda}{\text{\usefont{OML}{txmi}{m}{it}\symbol{"15}}}

\shorttitle{Data-driven model  of pipe flow}
\shortauthor{C. R. Constante-Amores, A. J. Linot and M. D. Graham}

\title{Dynamics of a Data-Driven Low-Dimensional Model of Turbulent Minimal Pipe Flow}

\author{C. Ricardo Constante-Amores$^{1,2}$\corresp{\email{crconsta@illinois.edu}}, Alec J. Linot$^{3}$, Michael D. Graham$^{1}$ % 
}
  
\affiliation{
$^1$Department of Chemical and Biological Engineering, University of Wisconsin-Madison, Madison WI
53706, USA
\\[\affilskip]
$^2$Department of Mechanical Science and Engineering, University of Illinois, Urbana Champaign, IL 61801, USA
\\[\affilskip]
$^3$Department of Mechanical and Aerospace Engineering, University of California, Los Angeles, CA 90095, USA
}
\begin{document}

\maketitle

\begin{abstract}
The simulation of turbulent flow requires many degrees of freedom to resolve all the relevant time and length scales. However, due to the dissipative nature of the Navier-Stokes equations, the long-term dynamics are expected to lie on a finite-dimensional invariant manifold with fewer degrees of freedom. In this study, we build low-dimensional data-driven models of pressure-driven flow through a circular pipe. We impose the `shift-and-reflect' symmetry to study the system in a minimal computational cell (e.g., the smallest domain size that sustains turbulence) at a Reynolds number of 2500. We build these models by using autoencoders to parametrize the manifold coordinates and neural ODEs to describe their time evolution. Direct numerical simulations (DNS)  typically require on the order of $\mathcal{O}(10^5)$ degrees of freedom, while our data-driven framework enables the construction of models with fewer than 20 degrees of freedom. Remarkably, these reduced order models effectively capture crucial features of the flow, including the streak breakdown. In short-time tracking, these models accurately track the true trajectory for one Lyapunov time, \textcolor{black}{as well as the leading Lyapunov exponent,}
while at long-times, they successfully capture key aspects of the dynamics such as Reynolds stresses and energy balance. \textcolor{black}{The model can quantitatively capture key characteristics of the flow, including the streak breakdown and regeneration cycle.}
Additionally, we report new  exact coherent states (ECS)  found in the  DNS with the aid of these low-dimensional models. 
This approach leads to the discovery of seventeen previously unknown solutions within the turbulent pipe flow system, notably featuring relative periodic orbits characterized by the longest reported periods for such flow conditions.
\end{abstract}

\section{Introduction}

The prevalence of wall turbulence in a diverse range of industrial and everyday applications has attracted substantial interest, as 
approximately $25\%$ of the energy consumed by industry is dedicated to transporting fluids through pipes and channels, and about one-quarter of this energy is dissipated due to turbulence occurring near walls \citep{jimenez_1991,Jimenez_2013,Avila}.  Pipe flow has been the subject of extensive research since the groundbreaking experiments conducted by Reynolds nearly two centuries ago \citep{Reynolds}. 
\CRCAr{Flow regimes are solely governed by the Reynolds number, which characterizes the ratio of inertial forces to viscous forces. Studies of the transition of pipe flow from laminar to turbulent   have resulted in numerous comprehensive reviews}
(see e.g. \citet{Mullin, Eckhardt, Smits, Avila}).

 The focus of the present work is the extent to which modern data-driven methods can capture the nonlinear dynamics of turbulent pipe flows near transition \CRCAr{\citep{darbyshire_mullin_1995,eckhardt2009introduction}}.
Because of its geometric and dynamical simplicity, the ‘minimal flow unit’ (MFU) of pipe flow has been previously presented  \citep{willis_avila_2013, Willis_2016, budanur_jfm_2017, Kaszas_Haller_2024,ecs_koopman}. The MFU represents the smallest domain size capable of sustaining turbulence, described by \citet{jimenez_1991} in the context of plane Couette flow. It encapsulates the essential elements of turbulent dynamics, particularly the `self-sustaining process' (SSP) described by \citet{hamilton_kim_waleffe_1995}. In this process, low-speed streaks near the wall become unstable and wavy, leading to their breakdown and the formation of rolls. These rolls then lift fluid from the walls, thereby regenerating the streaks and perpetuating the cycle.

\CRCA{To understand the intricate nonlinear dynamics of turbulence, researchers have adopted a dynamical systems perspective. The turbulent nonlinear dynamics of fluids are governed by the (infinite-dimensional) Navier-Stokes equations (NSE). Despite this infinite-dimensionality, the long-time dynamics
are expected to lie on a finite-dimensional invariant manifold within state space \CRCAr{\citep{Hopf_1948,Temam,cvitanovic2005chaos}} (we discuss this point in more detail below). From this viewpoint, turbulence can be seen as a chaotic attractor of the NSE. Turbulent flows \CRCAr{can display} persistent patterns in space and time, commonly known as exact coherent states (ECS) \citep{ Kawahara_arfm,Graham_ecs}.  There are several ECS types: steady or equilibrium solutions, periodic orbits, travelling waves, and relative periodic orbits. 
\CRCAr{A trajectory on the attractor picks up characteristics of underlying unstable ECSs as it approaches them but is ultimately moved away along unstable manifolds. When many of these ECSs are characterized, they can be used to approximate the statistical properties of the turbulent attractor, such as work by  \citet{nagata_1990,KAWAHARA_KIDA_2001,chandler_kerswell_2013,Page_2024}}.
% It has been proposed that the turbulent attractor can be decomposed into a collection of such ECS and by modelling the time spent by solutions near each ECS and transitions between them, the long-time statistics of the turbulent attractor can be predicted
However, fixed-point ECS cannot capture the dynamics entirely; periodic orbits can represent key aspects of nonlinear turbulent dynamics, such as bursting behavior \CRCAr{\citep{Cvitanovic_2013}}. In the realm of pipe flow, early studies focused on ECS and their role in the transition to turbulence \citep{Faisst_Eckhardt, wedin_kerswell_2004, Pringle_2007, Pringle_2009, DUGUET_WILLIS_KERSWELL_2008, Willis_2008, viswanath2009stable}. The first set of ECS discovered, presented by \citet{Faisst_Eckhardt, wedin_kerswell_2004}, were traveling-wave solutions, which were also observed in experimentally \citep{Hof_2004,de_Lozar}. To date, several studies  have focused on the discovery and classification of ECS according to the value of their average dissipation,  wave speed, and spatial symmetry \citep{Pringle_2009,willis_avila_2013, budanur_jfm_2017}.\citet{willis_avila_2013} and \citet{Willis_2016} reported 29 solutions and visualized these solutions in state space using symmetry reduction, showing the connections of relative periodic orbits and the turbulent attractor. \citet{budanur_jfm_2017} reported  49 new  relative periodic orbits and 10 traveling wave solutions. Their findings further supported the view    that turbulence wanders around ECS.
}

% The study of turbulence from a dynamical system point of view helps the understanding of its transition to turbulence \citep{Avila,wedin_kerswell_2004,nagata_1990, Gibson_2008,Gibson_2009,Halcrow_2009}. In this context, turbulent  flows can be viewed as a chaotic saddle \st{ a strange repeller} within the infinite-dimensional state-space  defined by the solutions to the Navier-Stokes equations (NSE) \citep{Graham_ecs} \MDG{we never said anything in the review about turbulence being a repeller}. Then,  fully developed turbulence can be seen as a state space populated with `invariant solutions' or `exact coherent states' (ECS)  whose stable and unstable manifolds guides the turbulent attractor \MDG{so is turbulence an attractor or a repeller?}. Thus, if many of these ECS are captured and characterised, their average  can reproduce the  statistics of the turbulent attractor \citep{Page_2024}. This idea has driven a large scientific interest in finding ECS for Kolmogorov, Couette  flow and pipe flow \citep{nagata_1990,wedin_kerswell_2004,waleffe_2001, Graham,page_kerswell_2020}.

% These numerical solvers can be used to solve the NSE in the search  of ECS embedded in the turbulent attractor.
However, identifying ECS remains challenging due to the high dimensionality of the state space. Traditional Newton-Raphson methods can be employed to locate those solutions, but more advanced techniques, such as Jacobian-Free Newton-Krylov method are more effective because they avoid explicit computation and inversion of Jacobian, which is expensive for high dimensional systems. In the latter,  the Jacobian matrix is not \CRCAr{explicitly} calculated as \CRCAr{detailed in \citet{Viswanath_2007}}. Good initial conditions are important 
%to ensure the method's success due to several factors: the high dimensionality of the systems; the nonlinearity of the NSE; and the optimization landscape associated with finding ECS is 
due to nonconvexity mostly and also computational expensive.
% often nonconvex, containing many local minima and saddle points (e.g., Newton-based methods tend to converge to local minima).
Favorable initial conditions help narrow the search area and increase the likelihood of finding ECS.
\CRCAr{\citet{Lan_Cvitanovi} proposed a variational method to find unstable periodic orbits in Kolmogorov flow, demonstrating that their method can converge to a broader set of solutions  compared to traditional  shooting methods.}

\CRCAr{A promising avenue for studying turbulence is the development of reduced-order models, which simplify the complex dynamics of turbulent flows while retaining essential features. } Among the nonlinear approaches to model reduction, invariant-manifold-based frameworks-particularly spectral submanifold (SSM) methods-have emerged as powerful tools.
% Among the nonlinear frameworks  such  invariant-manifold-based approaches and  spectral submanifolds methods (SSM),
The SSM  facilitates the construction of invariant manifolds near known stationary points, \CRCAr{ to which the dynamics of a system can be reduced \citep{li2022nonlinear,kogelbauer2018rigorous}.  SSMs represent the smoothest nonlinear extensions of the spectral subspaces of the linearized system near a stationary state, such as a fixed point or a periodic orbit.}
Recently, \citet{Kaszas_Haller_2024} employed SSM to successfully identify the invariant manifold \CRCAr{capturing} the edge of chaos in pipe flow. This manifold serves as a crucial boundary, demarcating the transition from the laminar state to turbulence within the phase space of the NSE. \CRCAr{We note that SSM  must be `anchored' to a known stationary point.}  In contrast, the framework we adopt in this work does not rely on such anchoring, allowing for a more flexible exploration of the turbulent attractor.  

The accurate simulation of MFU pipe flow requires a large state space to resolve all the relevant spatial and temporal scales. For instance, \citet{willis_avila_2013} and \citet{budanur_jfm_2017} required on the order of $\mathcal{O}(10^5)$ degree of freedom to capture the complex, nonlinear turbulent dynamics.
Performing data-driven modeling in this full state space is computationally challenging. However, due to the dissipative nature of the NSE, it is expected that viscosity attenuates the high wavenumber modes, confining the long-term dynamics to an invariant manifold with fewer degrees of freedom than the full state dimension \citep{Temam,Zelik}. The exact dimension of this invariant manifold is not known beforehand and must be estimated from data. The most common method for linear dimension reduction is principal component analysis (PCA),  also known as proper orthogonal decomposition (POD) in the fluid dynamics community. PCA works by projecting the state onto the set of orthogonal modes that capture the maximum variance or energy in the data \citep{Jolliffe1986, abdi2010principal,Holmes_2012}. However, PCA assumes a flat manifold because it is an inherently linear technique, which makes it a poor approximation for complex nonlinear problems. To address this,  nonlinear techniques for dimension reduction have emerged such as autoencoders. Autoencoders consist of a pair of neural networks in which one network maps from a high-dimensional space to a low-dimensional space, and the other maps back \citep{Kramer,Hinton,Milano}. For very high-dimensional systems, it can be advantageous to perform an initial linear dimension reduction step with PCA, followed by further nonlinear dimension reduction using an autoencoder \citep{linot_Couette,Young_simple,koopman_Couette}. Additionally, combining PCA and an autoencoder in parallel allows for capturing both linear and nonlinear features of the data, with the autoencoder refining the representation beyond what PCA alone can provide  \citep{alec_pre}.

Once we have a low-dimensional representation of the full-state, we can proceed in data-driven modelling of the dynamics in manifold coordinates \CRCAr{(i.e., the intrinsic variables that describe the key behavior of the system in the low-dimensional representation)}.  The goal is to learn a vector field that governs the evolution of the system in this low-dimensional representation. This approach has been successfully applied to chaotic systems, including the 1D Kuramoto-Sivashinsky equation \citep{alec_pre,ssm_chaos}, 2D Kolmogorov flow \citep{carlos_prf}, Couette flow \citep{linot_Couette,Kaszas}. \citet{alec_pre} presented the framework known as DManD which stands for `data-driven manifold dynamics' (DManD). In DManD, an autoencoder finds a low-dimensional representation of the full state, and then a neural ODE (NODE)  learns an evolution equation of this low-dimensional representation. \CRCAr{NODE is  a neural network that parameterises  the vector field of the latent space (e.g., low-dimensional coordinate representation found by the autoencoder)} \citep{chen2019neural,alec_chaos}. It is important to highlight that DManD is highly
 advantageous because, like the underlying turbulent systems, it is Markovian \CRCAr{in} nature  (where predictions of the next state only depend on the current state) and continuous-time formulation.

In this work, we address data-driven modeling for turbulent MFU pipe flow at $Re=2500$. 
\CRCAr{We note that while our approach shares methodological similarities with the recent work of      \citet{linot_Couette} on Couette flow, specifically the use of POD, autoencoders, and NODE, but the focus of the present study is on pipe flow, which poses fundamentally different physical challenges. Unlike the planar, zero-mean shear profile of Couette flow, pipe flow features a non-zero, radially varying mean velocity and geometric curvature, resulting in richer dynamics and more intricate turbulent structures. This work thus applies manifold-based data-driven modeling techniques to a more practically relevant and dynamically complex shear flow system. We show that the essential dynamics of pipe flow evolve on a low-dimensional manifold, enabling accurate reconstruction of both short-time trajectory evolution and long-time statistical properties.
We compute the Lyapunov spectrum on the manifold and compare the leading Lyapunov exponent with that obtaine from the DNS. The good agreement indicates that the model successfully captures the dominant dynamics, suggesting that only a few degrees of freedom are required. In addition, we identify ECS in the latent space and successfully converged them in the DNS, leading to the discovery of previously unreported solutions in this flow configuration.   We also acknowledge that in \citet{ecs_koopman}, the authors constructed data-driven models using pipe flow data restricted to a single relative periodic orbit, whereas the present study focuses on learning a low-dimensional model from trajectories embedded within the full turbulent attractor, leading to a much more general data-driven model (which is needed to discover new ECS). }
The rest of this paper is structured as follows. In section \ref{framework_section}, we 
describe the framework for dimension reduction and time evolution. In section \ref{results}, we present the results that include the dimension reduction and the predictions of the DManD model for short- and long-time statistic, ECS identification, and new ECS found in the DNS using  converged ECS from the model as initial conditions. Finally, in Section \ref{conclusion}, we summarize the concluding results.

\section{Framework \label{framework_section}}

\subsection{Dimension reduction}

\CRCA{While the state-space of a PDE is formally infinite-dimensional, the Navier-Stokes equations, which govern the motion of fluids, are dissipative in nature, and therefore solutions are expected to converge to a finite-dimensional invariant manifold, denoted as $\mathcal{M}$ in this context}  \citep{Temam,Foias,Zelik,Hopf_1948}. This manifold $\mathcal{M}$ exhibits a local Euclidean structure, implying that each point within it possesses a nearby region that can be bi-directionally mapped to and from a Euclidean space denoted as $\mathbb{R}^{d_\mathcal{M}}$, where $d_\mathcal{M}$ 
(with $d_\mathcal{M} \leq d_h$) 
represents the dimension of the manifold, in this work $d_h$ is a higher dimension, in which the manifold can be embedded.  This fact \CRCAr{is} also what allows for the global coordinate representation, since dynamics are learned in $d_h$ rather than $d_\mathcal{M}$. To effectively characterize this manifold and consequently understand the underlying system dynamics, only $d_\mathcal{M}$ independent coordinates are necessary, at least within local contexts. As $\mathcal{M}$ remains unchanged by the system dynamics, the vector field \CRCAr{which describes the dynamics} on $\mathcal{M}$ is always tangential to the manifold, 
%leading to Markovian dynamics on $\mathcal{M}$.
\CRCAr{resulting in deterministic, memoryless dynamics confined to $\mathcal{M}$}. Then, these dynamics are governed by an ordinary differential equation defined by this tangential vector field. 

\textcolor{black}{
In this work, there are four distinct representations of the system state. Let $\mathcal{H}$ denote the infinite-dimensional solution space of the NSE.
The direct numerical simulation (DNS) produces trajectories in a finite-dimensional subspace $\mathbb{R}^{d} \subset \mathcal{H}$, which we refer to as the `full state'. This full state is projected onto a $d_{\mathrm{POD}}$-dimensional subspace via POD, yielding a linear mapping $\mathbf{P} : \mathbb{R}^{d} \to \mathbb{R}^{d_{\mathrm{POD}}}$. The POD reduced representation is then mapped to a $d_h$-dimensional coordinate system via a nonlinear mapping $\mathcal{E} : \mathbb{R}^{d_{\mathrm{POD}}} \to \mathbb{R}^{d_h}$, obtained from a trained autoencoder.}

We consider a system that is characterized by deterministic, Markovian dynamics, so if $\boldsymbol{u} \in \mathbb{R}^{d}$ represents the full space state, then the dynamics can be represented by an ODE as
\begin{equation} \label{eq:ODE}
	\dfrac{d\boldsymbol{u}}{dt}=\boldsymbol{f}(\boldsymbol{u}).
\end{equation} 
here, $\boldsymbol{u}$ represents the full state-space. In practice, \CRCAr{$\boldsymbol{u}$} is obtained from DNS. \CRCAr{In this work,  we  find a mapping to a lower-dimensional representation } 

\begin{equation} \label{}
	\boldsymbol{h}=\boldsymbol{\chi}(\boldsymbol{u}),
\end{equation}

\noindent
where $\boldsymbol{h}\in\mathbb{R}^{d_h}$ is the low-dimensional representation of the full-state space, along with an approximation of its inverse 
\begin{equation} \label{eq:chunk2}
	\tilde{\boldsymbol{u}}=\check{\boldsymbol{\chi}}(\boldsymbol{h}),
\end{equation}
so that the full state space may be recovered  (e.g., ideally $\boldsymbol{u} \approx \tilde{\boldsymbol{u}}$).
% Once a small representation of the  dynamics  has been discovered, the time evolution is done  according to 
% \begin{equation} \label{eq:chunk3a}
% 	\dfrac{d\boldsymbol{h}}{dt}=g(\boldsymbol{h}).
% \end{equation} 
In this work, we opt to parameterise  $\boldsymbol{\chi}$, $\check{\boldsymbol{\chi}}$ with an autoencoder, referred to as a hybrid autoencoder in \cite{alec_pre}. This hybrid autoencoder is based on the idea of using neural networks to learn the corrections from the leading POD coefficients

\begin{equation}
\boldsymbol{h}=\boldsymbol{\chi}(\boldsymbol{u};\theta_\mathcal{E})=\mathsfbi{U}^T_{d_h} \boldsymbol{u}+\boldsymbol{\mathcal{E}}(\mathsfbi{U}^T_{d_{POD}}\boldsymbol{u},\theta_\mathcal{E}),
\end{equation}

\noindent
here, $\mathsfbi{U}_k \in \mathbb{R}^{d\times k}$ corresponds to a matrix containing the first $k$ POD modes  ordered by variance, and $\boldsymbol{\mathcal{E}}$ corresponds to the encoder in the neural network \CRCAr{(e.g, section 3.2.1 presents the framework for the linear reduction with POD)}. In this way, the first term ($\mathsfbi{U}^T_{d_h}$) is the  projection onto the leading $d_h$ POD modes,  and the second term is the corrections provided by the neural network. The mapping back to the full space is given by

\begin{equation}
\tilde{\boldsymbol{u}}=\check{\boldsymbol{\chi}}({\boldsymbol{h};\theta_\mathcal{E}})=\mathsfbi{U}_{d_{POD}} ([\boldsymbol{h},0]^T+\boldsymbol{\mathcal{D}}(\boldsymbol{h};\theta_\mathcal{D})).
\end{equation}

\noindent
here, $[h, 0]^T$ is the $\boldsymbol{h}$ vector zero-padded to the correct size, and $\mathcal{D}$ is a neural network. The first term is the POD mapping back to the full space, and the second term is a NN correction. The weight parameters $\theta_\mathcal{E}$, $\theta_\mathcal{D}$ are trained to minimize the loss

\begin{equation} \label{loss_auto}
	L=\dfrac{1}{dK}\sum_{i=1}^K||
 \boldsymbol{u}(t_i)-\check{\boldsymbol{\chi}}(\boldsymbol{\chi}(\boldsymbol{u}(t_i);\theta_\mathcal{E});\theta_\mathcal{D})||^2 +
 \dfrac{1}{d_hK}\sum_{i=1}^K \kappa||
 \boldsymbol{\mathcal{E}}(\mathsfbi{U}^T_{r}\boldsymbol{u}(t_i);\theta_\mathcal{E})+\boldsymbol{\mathcal{D}}_{d_h}(\boldsymbol{h}(t_i);\theta_\mathcal{D})
 ||^2 
\end{equation}
here, the first term corresponds to the mean-squared error of the reconstruction $\tilde{\boldsymbol{u}}$, while the second term corresponds to a penalty term to enhance the accurate representation of the leading $d_h$ POD coefficients (e.g., $\boldsymbol{\mathcal{D}}_{d_h}$, denotes the leading dh elements of the decoder output). Here $\kappa$ is a a penalty term.
This penalty does not directly reduce the magnitude of the encoder's correction; instead, it promotes its removal by the decoder. Throughout, the norm is defined as the $L_2$ norm, $\|\boldsymbol{q}\|^2$. The prefactor in front of each term accounts for averaging over the vector components and the batch size $K$. In section \ref{manifold_coord}, we also  use  standard autoencoders, which can be seen as $\boldsymbol{h}=\boldsymbol{\chi}(\boldsymbol{u};\theta_\mathcal{E})=\boldsymbol{\mathcal{E}}(\mathsfbi{U}^T_{POD}\boldsymbol{u},\theta_\mathcal{E})$ and $\tilde{\boldsymbol{u}}=\check{\boldsymbol{\chi}}(\boldsymbol{h};\theta_\mathcal{E})=\mathsfbi{U}^T_{POD}\boldsymbol{\mathcal{D}}(\boldsymbol{h};\theta_\mathcal{D})$, to highlight the effectiveness of using the hybrid autoencoders to find an accurate representation of the manifold coordinates. We note that this \CRCAr{hybrid} autoencoder has been used successfully for the Kuramoto-Sivashinsky equation, chaotic Kolmogorov flow, and MFU plane Couette flow \citep{alec_pre,linot_Couette,carlos_prf}.

To train both the hybrid and standard autoencoders, we use the Adam optimizer to minimize the loss function presented in Equation \ref{loss_auto}, utilizing the  POD coefficients as inputs (as explained in section \ref{sec_pod}). The training process spans 500 epochs, and we incorporate a learning rate scheduler that reduces the learning rate from $10^{-3}$ to $10^{-4}$ after the initial 300 epochs. This adjustment is made based on our observation that no significant improvements in reconstruction error occur beyond this number of epochs. For the hybrid autoencoder approach, we set the hyperparameter $\kappa= 0.1$, while for the standard autoencoder, $\kappa=0$  (indicating that this term is not included). \textcolor{black}{All relevant details of the neural network architectures and their hyperparameters (e.g., number of layers, neurons per layer, activation functions) are summarized in Table 1, to ensure reproducibility of the results.} The specific network parameters were determined through a meticulous \CRCAr{trial and error} search, exploring variations in the network's architecture and activation functions. We remark that our goal was to achieve the lowest reconstruction error while avoiding excessive computational costs.

\subsection{Time evolution:  neural ODEs}

\CRCA{We use a stabilized neural ODE framework for state-space modeling in the latent space.
Rather than equation 2.1, we use a slight modification:
}

\begin{equation} \label{eq:Node1}
	\dfrac{d\boldsymbol{h}}{dt}=\boldsymbol{g}(\boldsymbol{h})-\boldsymbol{A}\boldsymbol{h}.
\end{equation} 
here $\boldsymbol{A}=\gamma \delta_{ij}\sigma_i(\boldsymbol{h})$,
% represents the stabilization term that can be learned from data or fix it. Here, we opt for the latter, and we set it to be the diagonal matrix
where $\sigma_i(\boldsymbol{h})$ stands for the standard deviation of the $i$th component of $\boldsymbol{h}$, $\gamma$ is a fixed parameter and $\delta_{ij}$ is the Kronecker delta. This modification, with a small value of $\gamma$, has been found to stabilize the dynamics against spurious growth of fluctuations without compromising the accuracy of predictions \citep{linot_Couette,alec_jcp}.
\CRCAr{\citet{linot_Couette} demonstrated the importance of this damping term in detail for MFU plane Couette flow.}
% The rationale behind introducing linear damping lies in the behavior of dissipative systems, where viscosity dampens the high wave numbers. In such systems, data located on the attractor exhibit a minimal content in these high wave numbers, and the incorporation of linear damping becomes crucial as it enhances the stability of the trajectories and also aligns with the observed dynamics. 

\CRCAr{Next, we approximate {\bf g} using a neural ODE. For training {\bf g},}
we integrate  equation \ref{eq:Node1} forward in time from $t_i$ to $t_i + \tau$ resulting in the prediction

\begin{equation}\label{eq:Node2}
	\tilde{\boldsymbol{h}}(t_i+\tau)=\boldsymbol{h}(t_i)+\int_{t_i}^{t_i+\tau}\boldsymbol{g}(\boldsymbol{h}(t);\theta_g)-\boldsymbol{A}\boldsymbol{h}(t)dt,
\end{equation}

We determine the parameters $\theta_g$ by minimising the difference between the true state $\boldsymbol{h}(t_i+\tau$) and the  predicted state $\tilde{\boldsymbol{h}}(t_i+\tau)$, as 

\begin{equation} \label{eq:Node_loss}
	J=\dfrac{1}{d_hK}\sum_{i=1}^K||\boldsymbol{h}(t_i+\tau)-\tilde{\boldsymbol{h}}(t_i+\tau)||^2.
\end{equation}

To calculate the derivatives of $\boldsymbol{g}$ with respect to the \CRCAr{neural networks parameters $\theta_g$}, we make use of automatic differentiation.
\CRCAr{We train the stabilized NODE to predict the system evolution over one time unit, using data from which the temporal mean has been subtracted,} by employing the Adam optimizer in PyTorch to minimize the loss function described in equation \ref{eq:Node_loss} \citep{pythorch}. 
The training process incorporates a learning rate scheduler, which decreases the learning rate at three evenly spaced intervals, continuing until the learning curve stabilizes. The specific details of this neural network are provided in Table \ref{table_arch}. The architectural choices are empirical, determined through trial and error by varying the number of nodes and layers.

Once $\boldsymbol{g}$ is determined, an arbitrary initial condition can be mapped into the low-dimensional coordinates with  $\boldsymbol{\chi}$. Then, the state evolution of $\boldsymbol{h}$ at arbitrary points in time can be computed as the solution to equation \ref{eq:Node1}, and finally the solution can be mapped back to the full space with $\check{\boldsymbol{\chi}}$.

\begin{figure}
\centering
\begin{tabular}{cc}
\includegraphics[width=0.45\textwidth]{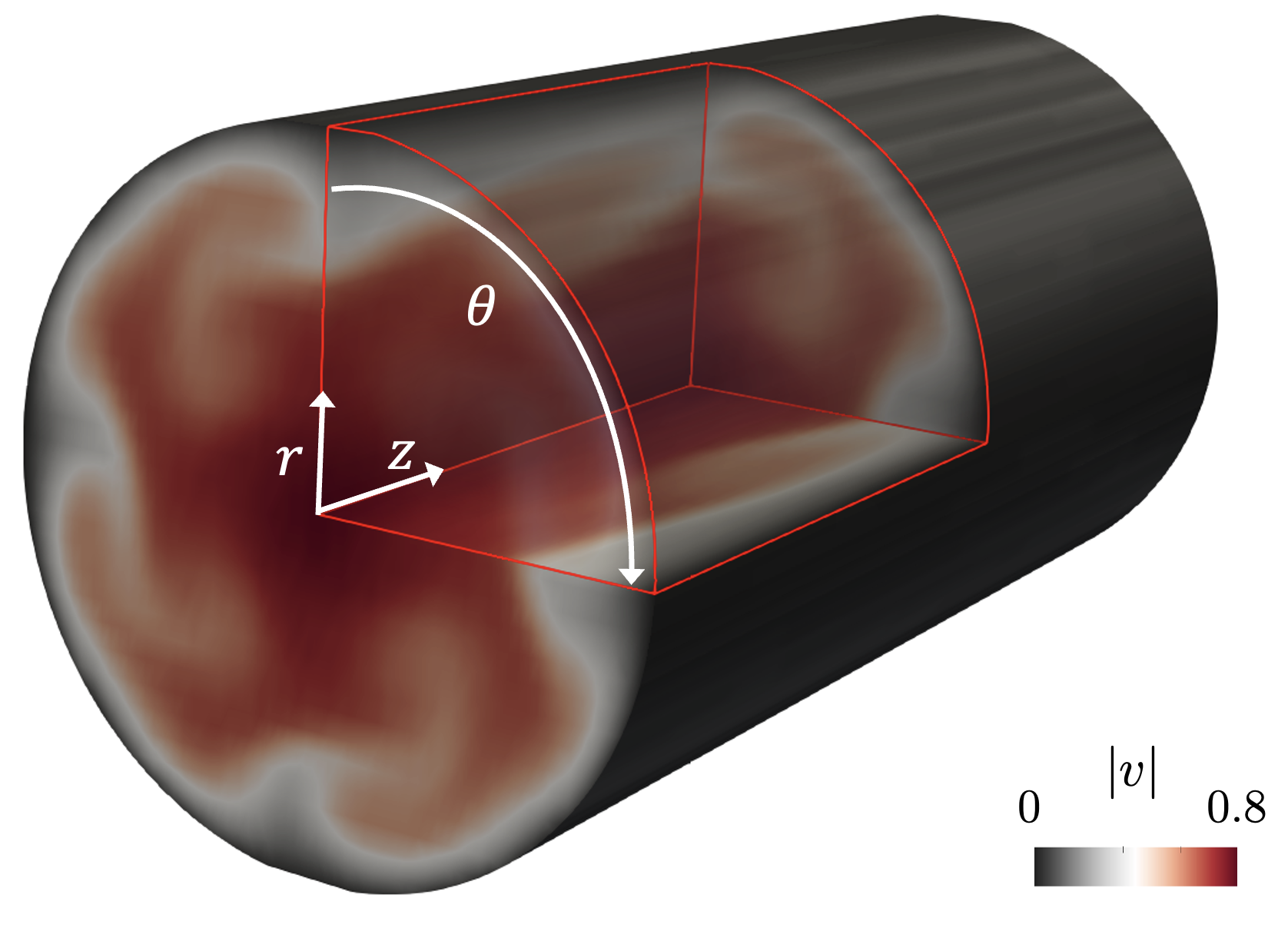}&
\includegraphics[width=0.45\textwidth]{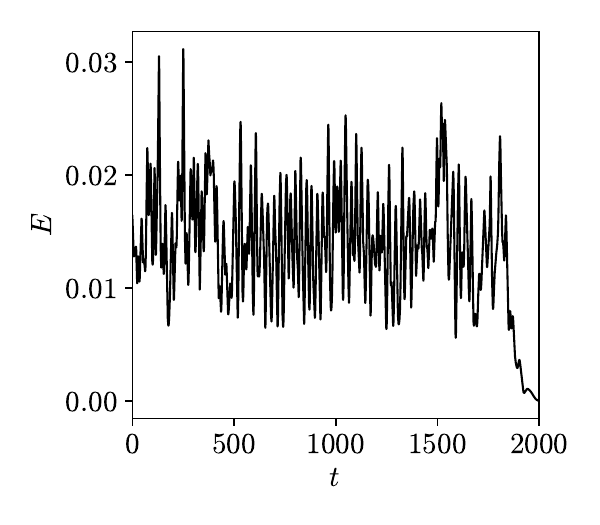}\\
(a) & (b) \\
\end{tabular}
\caption{Schematic representation of the three-dimensional pipe flow system. Panel (a) shows a snapshot of the magnitude of the velocity field. For visualisation purposes, the entire pipe is shown, although calculations in this work is restricted in the shift-and-reflect symmetry subspace with $m_p=4$ (\CRCAr{ whose boundaries are highlighted with solid red lines}). Panel (b) represents the energy in the axially dependent modes only (k non-zero). This quantity decays rapidly after relaminarisations. \label{framework}}
\end{figure}

\section{Results \label{results}}

In this section, we provide a detailed description of the dataset for MFU pipe flow, present the results of dimensionality reduction,  evaluate the performance of the reduced models as we vary their dimension, and introduce new ECS. Figure \ref{framework}a shows a three-dimensional representation of the MFU pipe configuration used in the current research. For visualisation purposes, the entire pipe is shown, although calculations in this work are confined to the shift-and-reflect symmetry subspace  (which is the highlighted area with opacity).

\subsection{Description of pipe flow data}

We perform DNS of an incompressible viscous fluid moving inside of a pipe with a circular cross-section. We consider flow with a constant max flux, thus, the dimensionless forms of the Navier-Stokes equations are expressed as

\begin{equation*}
\frac{\partial \boldsymbol{v}}{\partial t}+
\boldsymbol{U} \cdot\nabla \boldsymbol{v} + 
\boldsymbol{v} \cdot\nabla \boldsymbol{U} +
\boldsymbol{v} \cdot\nabla \boldsymbol{v} 
= -\nabla p
+ 32 \frac{\beta}{\Rey} \boldsymbol{z}
+\frac{1}{Re} \nabla^2 \boldsymbol{v},
\end{equation*}
\begin{equation}\label{div}
 \nabla \cdot \boldsymbol{v}=0.
\end{equation}
The equations are solved in cylindrical coordinates 
$(r,\theta, z)$ which refer to  radial, azimuthal and the streamwise (axial) directions, respectively. 
Here, $\boldsymbol{v}$ and $p$ stand for the velocity and the pressure, respectively.  
The Reynolds number $\Rey$  is defined as $\Rey=U\mathscr{D}/\nu$, where U, $\mathscr{D}$ and $\nu$ are the mean flow velocity,  
the pipe diameter and the kinematic viscosity, respectively.
Lengths and velocities are made non-dimensional using  $\mathscr{D}$ and $U$ as characteristic values, and hence, time will be made non-dimensional using $\mathscr{D}/U$. The velocity $\boldsymbol{v}=(v_r, v_\theta,v_z)$ represents the deviation from laminar Hagen-Poiseuille flow equilibrium $\textbf{U}(r)=2 (1-(2r)^2)\textbf{z}$. To maintain constant mass flux, a pressure gradient is required, and the excess pressure needed is measured by the feedback variable $\beta = \beta (\boldsymbol{v})$; thus the total dimensionless pressure gradient is $(1 + \beta)(32/Re)$, and $\beta=0$ for laminar flow.

In the NSE, symmetries appear in the form of continuous and discrete symmetry groups.  For the former,  the cylindrical wall in pipe flow  limits rotational symmetry around the $z$-axis and restricts translational movement along it. 
Let $\CRCAr{g}(\phi, \ell)$ represent the shift operator, where $\CRCAr{g}(\phi, 0)$ signifies an azimuthal rotation by $\phi$ about the pipe axis, and $\CRCAr{g}(0, \ell)$ indicates the streamwise translation by $\ell$. Let $\sigma$ represent the reflection about the $\theta = 0$ azimuthal angle. Thus

\begin{equation*}
\left.\begin{matrix}
\CRCAr{g}(\phi, \ell)[v_r, v_\theta,v_z,p](r,\theta,z) = [v_r, v_\theta,v_z,p](r,\theta-\phi,z-\ell)
\\ 
\sigma[v_r, v_\theta,v_z,p](r,\theta,z) = [v_r, -v_\theta,v_z,p](r,-\theta,z)
\end{matrix}\right\}
\label{continuos_sym}
\end{equation*}

\noindent
Apart from azimuthal reflection, the NSE also have additional discrete symmetries in both azimuthal and streamwise directions across the computational cell $\Omega$.
The symmetry group of streamwise periodic pipe flow is $SO(2)_z \times O(2)_{\theta}$.
In this paper, we restrict the dynamics  to the 'shift-and-reflect' symmetry subspace

\begin{equation}
S = \{e, \sigma_\theta  \CRCAr{g}_{s} \},
\end{equation}
where $\CRCAr{g}_{s}$ represents a streamwise shift by $L/2$, i.e., flow fields of equation \ref{continuos_sym} that satisfy
\begin{equation}
[v_r, v_\theta,v_z,p](r,\theta,z)=[v_r, -v_\theta,v_z,p](r,-\theta,z-L/2).
\label{eq_SR}
\end{equation}

% To clarify, the shifting operation refers to the displacement of the flow pattern along the streamwise direction, while the  reflection operation, on the other hand, involves flipping the flow pattern across a plane perpendicular to $\theta=0$. This means that the flow pattern is mirrored, and the flow velocities are reversed.

This symmetry couples the streamwise translations with the azimuthal reflection. By imposing the shift-reflect symmetry, eliminates the continuous phase along the azimuthal rotations. In this work, we do not factor out the continuous symmetry in the streamwise direction. 
\CRCAr{Factoring out the streamwise symmetry reduces the manifold dimension by only one degree of freedom, and our focus is on developing a general framework that does not rely on symmetry reduction.}
% as in other work \citep{budanur_jfm_2017,Willis_2016}.

To perform the DNS  of the incompressible turbulent pipe flow under the assumption of shift-and-reflect symmetry, we use the pseudo-spectral code Openpipeflow \citep{willis2017openpipeflow}. 
Fourier discretization is used for the periodic axial ($z$) and azimuthal ($\theta$) directions, with $K$ and $M$ representing the number of Fourier modes, i.e.,

\begin{equation}
\label{openpipe_eq}
\left \{ \boldsymbol{v} \right \} (r,\theta,z,t)=\sum _{m=-M}^M \sum _{k=0}^K  \boldsymbol{B}_{mk}(r,t) e^{im_pm\theta} e^{ik\alpha z},
\end{equation}
here $\boldsymbol{B}_{mk}(r,t)$
represents a three-vector of Fourier amplitudes,  $m_p$ is a parameter to 
\CRCAr{control the azimuthal shift-reflect subspace we work in}, and $\alpha=2\pi/L$ is a parameter that controls the length of the pipe. In the radial direction, a Chebyshev grid is used that clusters points near the wall to effectively resolve the velocity gradients.
% In  physical space, the number of grid points in the $z$ and $\theta$ directions is increased by a factor of $3/2$ due to dealiasing. 
No-slip and no-penetration boundary conditions are enforced at the wall. For a more detailed description of the numerical method, we refer to \citet{willis2017openpipeflow}.

The simulation of the entire cross-sectional pipe ($m_p=1$) presents a naturally periodic azimuthal boundary condition, while other values of $m_p$ result in $\boldsymbol{v}$ repeating itself in the azimuthal direction. In this work, we construct models for MFU pipe flow at $Re=2500$, with $m_p=4$ (`shift-and-reflect' invariant subspace) and $\alpha=1.7$, as previous work done by \cite{Willis_2016} and \cite{budanur_jfm_2017}. Then,  the size of the computational cell   is described by

\begin{equation}
\Omega=[1/2,2\pi/m_p,\pi/\alpha]\equiv{(r,\theta,z)\in [0,1/2]\times[0,2\pi/m_p]\times [0,\pi/\alpha]}.
\end{equation}
% where $L=\pi/\alpha$ stands for the length of the pipe.
Thus, the  domain size   in wall units for the wall-normal, azimuthal, and streamwise dimensions is
 $\Omega^+ \approx [100,160,370]$, respectively, which compares well with the minimal flow units for Couette flow and plane Poiseuille flow (i.e., $\Omega^+ \sim[68,128,190]$ and $\Omega^+  \approx [>40,100,250-300]$, respectively). This domain size is similar to that used in the minimal box simulation by \citet{jimenez_1991} and \citet{willis_avila_2013}, and it is sufficiently large to exhibit complex chaotic behavior.

Data were generated with $\delta t=0.01$ on a grid $(N_r,M,K)=(64,10,14)$. \CRCAr{To eliminate aliasing errors in the evaluation of nonlinear terms, the $3/2$ rule is applied. This rule increases the number of grid points in each periodic direction by a factor of $3/2$, after converting the number of complex modes to the corresponding number of real physical grid points. Since each complex Fourier mode requires two real degrees of freedom in physical space, the number of grid points in each  direction becomes, $N_\theta=2M$ and  $N_z=2K$. Therefore, the velocity field is evaluated on a 
 $64\times30\times42$ grid in physical space, with three velocity components. The total number of degrees of freedom is: $d=N_r \times N_\theta \times N_z   \times  3= 241,920$, so $\boldsymbol{u} \in\mathbb{R}^{241,920}$.}
% Then, variables are evaluated on $64\times30\times42$ grid points,
% $\boldsymbol{u} \in\mathbb{R}^{241,920}$ \CRCAr{(i.e., $d=N_r \times 3M \times 3K   \times  3= 241,920$.
%where the factors of 3 account for the three velocity components and the 3/2 dealiasing rule applied in the axial and azimuthal directions
% )}. 
In this grid size, $(\Delta\theta D/2)^+ \approx 5.3$ and $\Delta z^+ \approx 8.8$ (which is consistent  with grid sizes used by  \citet{jimenez_1991}).  The resolution was tested to ensure mesh-independent results, confirmed by a drop in the energy spectra by at least 4 orders of magnitude.

We initiated simulations from random divergence-free initial conditions.
% The first initial condition comes from  a short-period RPO with a period $T=26.649$ for MFU pipe flow at $Re=2500$, $m_p=4$, $\alpha=1.7$, which is part of the database of Openpipeflow \citep{willis2017openpipeflow}.
The solutions were evolved forward in time for either 10,000 time units or until relaminarization occurred. The initial 100 time units were excluded as transient data, and the final 200 time units were excluded if relaminarization had occurred. 
% Figure \ref{framework}b plots the energy in the axially dependent modes only (k non-zero),  which decays rapidly after relaminarization, indicating the end of the simulation.
\CRCAr{Figure~\ref{framework}b shows the energy in the axially dependent modes only (i.e., modes with non-zero axial wavenumber \(k\)), which decays rapidly after relaminarisation, indicating the end of the simulation. Relaminarisation is identified by monitoring this energy, and following \citep{willis2017openpipeflow}, the simulation is terminated once it falls below \(10^{-5}\).}
This process was repeated with new initial conditions until we accumulated 96,921 time units of data, sampled at one-time unit intervals ($i.e.,\tau=1$). \CRCA{Consequently, all data lies on the turbulent attractor.} We divided this dataset, allocating $80\%$ for training and $20\%$ for testing purposes.

\textcolor{black}{ In terms of energy balance, the intermittent nature of relaminarization results in that the energy balance does not necessary hold true, especially when patching together trajectories from different simulations. While relaminarization events temporarily disrupt this balance, the entire  dataset is collected from regions where the flow remains on the turbulent attractor, where energy input and dissipation balance should hold. Thus, although the system is not strictly stationary due to relaminarization, all of the data ultimately represents dynamics within the chaotic saddle. This approach, while not ideal for strictly steady-state analysis, provides a robust basis for exploring turbulent dynamics across a variety of conditions. Future work to avoid this problem would  either increase the Reynolds number  or simulate the full pipe without symmetry restrictions.}

%--------------------------------------
\subsection{Learning of  manifold coordinates \label{manifold_coord}}
%--------------------------------------

In this section, we present our approach to dimensionality reduction. We first apply linear reduction using POD,  and then proceed with nonlinear reduction using autoencoders.

%--------------------------------------
\subsubsection{Linear dimension reduction with POD: From $\mathcal{O}(10^5)$ to $\mathcal{O}(10^3)$ \label{sec_pod}}
%--------------------------------------

\begin{figure}
\begin{center}
\begin{tabular}{cc}
\includegraphics[width=0.48\linewidth]{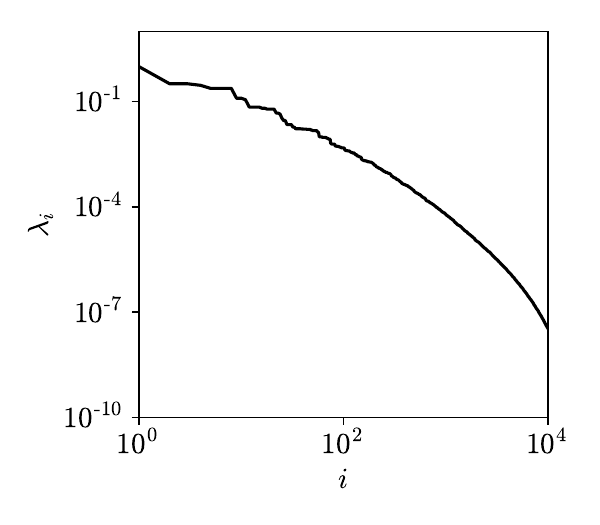}&
\includegraphics[width=0.48\linewidth]{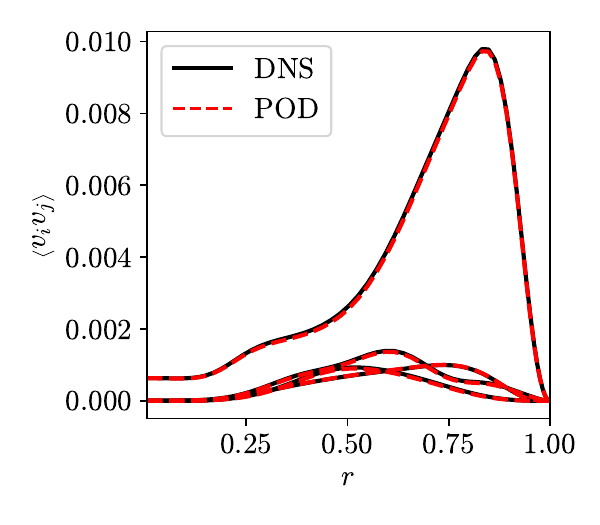}\\
(a) & (b)  \\
\end{tabular}
\end{center}
\caption{\label{POD}
(a) Eigenvalues of the POD modes sorted in descending order. (b) Reconstruction of four components of the Reynolds stresses from the DNS  and the data projected onto 512  POD  modes. The curves correspond to $ \left \langle u_z^2 \right \rangle, \left \langle u_\theta^2 \right \rangle, \left \langle u_r u_z \right \rangle$ and $\left \langle u_\theta^2 \right \rangle$, from top to bottom, respectively.
} 
\end{figure}

The first step in constructing the low-dimensional model is to apply  POD on the original dataset as a preprocessing step. This aims to reduce the dimension of the problem from approximately $\mathcal{O}(10^5)$ degrees of freedom to $\mathcal{O}(10^3)$, while preserving the essential characteristics of the turbulent flow system. POD tries to find the function $\boldsymbol{\Phi}$ that maximizes

\begin{equation}
    \frac{\left \langle \left |  (\boldsymbol{v'},\boldsymbol{\Phi})_E\right |^2  \right \rangle }{|| \boldsymbol{\Phi}  ||^2_E  },
\end{equation}

\noindent
where \CRCAr{$\boldsymbol{v'}(\boldsymbol{x})= \boldsymbol{v}(\boldsymbol{x}) - \bar{\boldsymbol{v}}(\boldsymbol{x}) $ is the fluctuating component of the velocity field, and $\bar{\boldsymbol{v}}$ is the mean velocity,  obtained by averaging over both space and time,  }
$\left \langle \cdot \right \rangle $ is the ensemble average, and the inner product is defined to be

\begin{equation}\label{eigenvalue1}
(\boldsymbol{q}_1,\boldsymbol{q}_2)_E= \int\int\int \boldsymbol{q}_1 \cdot \boldsymbol{q}_2 d\boldsymbol{x}
\end{equation}

\noindent
with the corresponding energy norm $  || \boldsymbol{q}  ||^2_E =(\boldsymbol{q},\boldsymbol{q})_E$. Solutions $\boldsymbol{\Phi}^{(n)}$ to this problem satisfy the eigenvalue problem
\begin{equation}\label{eigenvalue2}
%  \int_{0}^{L}\int_{0}^{2\pi/m_p} \int_{0}^{R}
% \left \langle
% \boldsymbol{v'}(\boldsymbol{x},t), \boldsymbol{v}^*(\boldsymbol{x'},t) \right \rangle 
% r'dr' d\theta' dz'=\lambda \boldsymbol{\Phi}^{(n)}, 
 \sum_{j=1}^3\int_{0}^{L}\int_{0}^{2\pi/m_p} \int_{0}^{R}
\left \langle
v'_i(\boldsymbol{x},t) v^{'}_j(\boldsymbol{x}',t) \right \rangle \Phi^{(n)}(\boldsymbol{x}') 
r'dr' d\theta' dz'=\lambda_i \Phi^{(n)}_i(\boldsymbol{x}), 
\end{equation}

% and $\boldsymbol{v}(\boldsymbol{x})= \boldsymbol{v'}(\boldsymbol{x}) - \bar{\boldsymbol{v}}(\boldsymbol{x}) $ stands for the fluctuating component of the velocity field, and $\bar{\boldsymbol{v}}$ is the mean velocity,  obtained by averaging over both space and time. 
The eigenvalue problem described by equation \ref{eigenvalue2} becomes $d\times d$. 
% We have already taken advantage of the discrete symmetry considerations  invariant solutions under the shift-and-reflect symmetry subspace.
To reduce the computational cost of this eigenvalue problem,  and preserve symmetries, we treat the POD modes as Fourier modes in both the azimuthal and streamwise directions. This approach has been previously applied by
\citet{pod_pipe} and \citet{linot_Couette} for turbulent pipe  and  plane Couette flow, respectively. 
\citet{Holmes_2012} showed that for translation-invariant directions, Fourier modes are the optimal POD modes.
Thus, the eigenvalue problem becomes

\begin{equation}
 % \sum_{j=1}^3 \int_{0}^{R} \left \langle \hat{v}^{'}_i (k_\theta,k_z;r) \hat{v}^{'*}_j (k_\theta,k_z;r') \right \rangle  \varphi^{*(n)}(k_\theta,k_z;r') r'dr' =\lambda ^{(n)}_{k_\theta k_z}\varphi^{(n)}(k_\theta,k_z;r) ,
 \sum_{j=1}^3 \int_{0}^{R} \left \langle \hat{v}'_i (r',k_\theta,k_z,t) \hat{v}'^{*}_j (r',k_\theta,k_z,t) \right \rangle  \varphi^{(n)}_{jk_\theta k_z}(r') r'dr' =\lambda ^{(n)}_{k_\theta k_z}\varphi^{(n)}_{ik_\theta k_z}(r).
\end{equation}
% where  $\lambda$ is the eigenvalue, and $\hat{\boldsymbol{v}} (k_\theta,k_z;r)$ is the Fourier transform of the fluctuating velocity in the azimuthal and axial direction.

\noindent
where \CRCAr{$*$ denotes the complex conjugate.} Thus, the eigenvalue problem is reduced from  $d\times d$ to a $3 N_r \times 3 N_r$ problem for each pair of wavenumbers $(k_\theta,k_z)$ in the Fourier coefficients. This analysis gives us POD modes represented by
\begin{equation}
\boldsymbol{\Phi}^{(n)}_{k_\theta k_z}(r,\theta,z)= \boldsymbol{\varphi} ^{(n)}_{k_\theta k_z}(r) e^{i k_\theta \theta} e^{i 2 \pi k_z z /L}.
\end{equation}
and eigenvalues $\lambda_{k_\theta k_z}^{(n)}$.
The projection onto these modes results in complex values unless both $k_{\theta}$ and ($k_{z}$ are zero). We arrange the modes in descending order of their eigenvalues ($\lambda_i$), and we select the leading 512 modes, resulting in a vector of POD coefficients (${\bf a}(t)$). Most of these modes are characterised by being complex-valued (i.e., they have 2 degrees of freedom), so projecting onto these modes results in a 1014-dimensional system, i.e., $d_{POD} = 1014$. In figure \ref{POD}a, we display the eigenvalues revealing a rapid decline followed by a long tail that contributes minimally to the energy content. By dividing the sum of the eigenvalues of the leading 512 modes by the total sum, we find that these modes account for  $99.44\%$ of the energy. Illustratively,  figure \ref{POD}b displays the reconstruction of Reynolds stresses for the components $ \left \langle v_z^2 \right \rangle, \left \langle v_\theta^2 \right \rangle, \left \langle v_r v_z \right \rangle$ and $\left \langle v_r^2 \right \rangle$, from top to bottom, respectively. This reconstruction is obtained  using those 512 modes with 5000 snapshots. We observe an excellent agreement between the DNS and the flow field obtained after truncating to the leading POD modes.

%--------------------------------------
\subsubsection{Nonlinear dimension reduction with autoencoders: From $\mathcal{O}(10^3)$ to $\mathcal{O}(10^1)$}
%--------------------------------------

\begin{table}
\centering
\begin{tabular}{cccc}
Function & Shape & Activation  & Learning Rate \\
\hline
 $ \boldsymbol{\chi}$		& 		1014/2500/1000/500/100/$d_h$ \quad           & ReLU/ReLU/ReLU/ReLU/lin         & $[10^{-3},10^{-4}]$ \\
  $\check{\boldsymbol{\chi}}$		& $d_h$/100/500/1000/2500/1014 \quad           & ReLU/ReLU/ReLU/ReLU/lin        & $[10^{-3},10^{-4}]$ \\
  $\boldsymbol{g}_{_{NN}}$	    &  $d_h$/250/250/250/250/$d_h$ \quad  & sig/sig/sig/sig/lin & $[10^{-2},10^{-3}]$ \\
\end{tabular}
\caption{Neural network architectures for the autoencoder and NODE. `Shape' represents the dimension of each layer, `Activation'  refers to the types of activation functions used, where  `ReLU', `sig' and `lin' stand for  ReLU, sigmoid and linear activation functions, respectively. `Learning Rate' represents the learning rate at various times during training.
\label{table_arch}}
\end{table}

\begin{figure}
\begin{center}
\begin{tabular}{cc}
\includegraphics[width=0.5\linewidth]{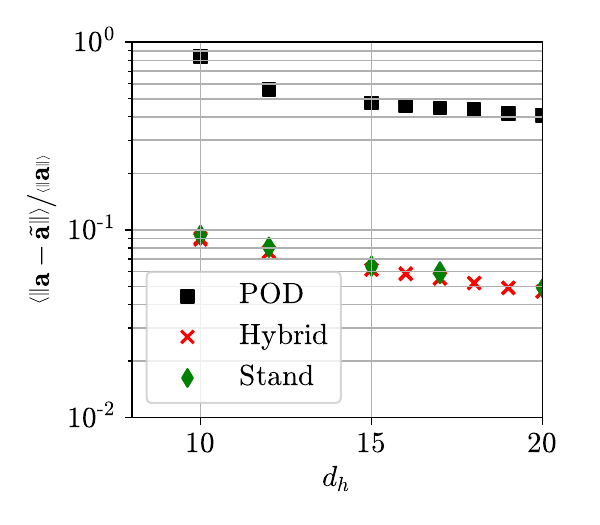}&
\includegraphics[width=0.5\linewidth]{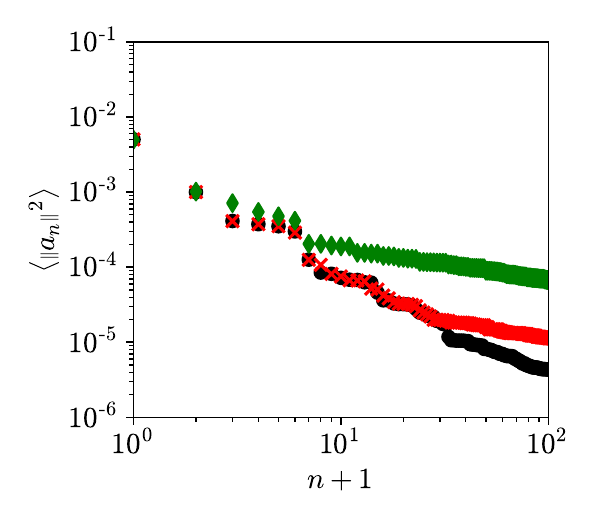}\\
(a) & (b)\\
\includegraphics[width=0.5\linewidth]{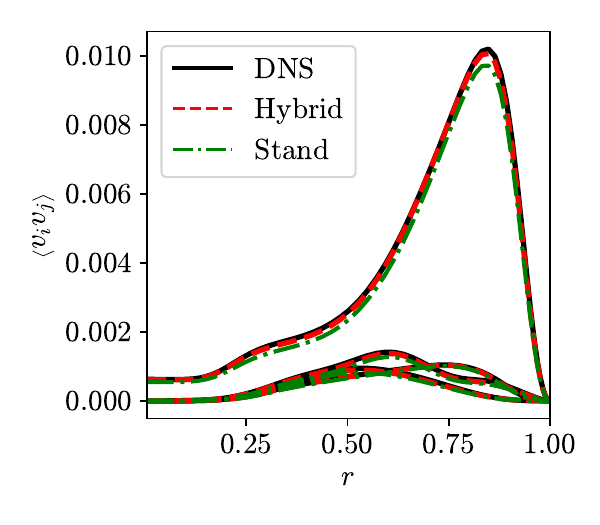} &
\includegraphics[width=0.5\linewidth]{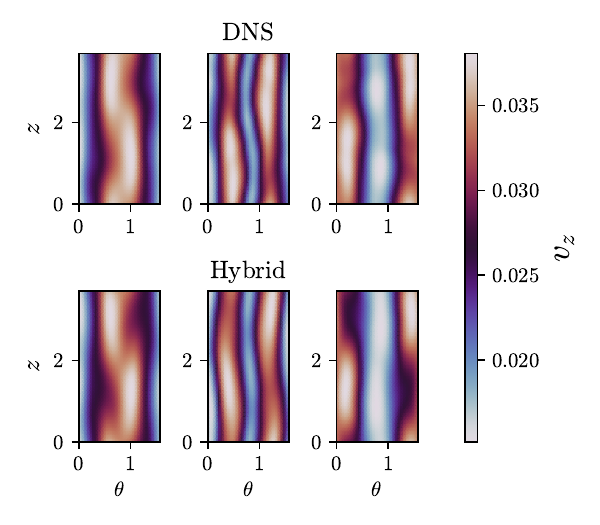} \\
\\
(c) & (d)\\
\end{tabular}
\end{center}
\caption{\label{auto_figure} Nonlinear reduction with autoencoders: (a) Relative error on test data for POD coefficients, standard and hybrid autoencoders as a function of the latent dimension $d_h$. For each dimension, results for two standard and two hybrid autoencoders are reported.
(b) Reconstruction of $\left \langle \left \| a_n \right \|^2 \right \rangle$ (mean-squared POD coefficient amplitudes) for the test data from 512 POD modes and the standard and hybrid autoencoders at $d_h = 20$.
(c) Components of the Reynolds stresses from the DNS  and using autoencoders with $d_h=20$.
(d) Two-dimensional representation of the flow field in a $z-\theta$ plane ($r = 0.496$) with $u_z$ for the DNS  and reconstructed using the hybrid autoencoder at $d_h = 20$.
} 
\end{figure}

After \CRCAr{projecting} the data to the leading POD modes,  and selectively retaining only the high-energy coefficients, our next step involves a nonlinear reduction of the data using autoencoders.  As a crucial preprocessing step, we normalize the POD coefficients by subtracting the mean and dividing by the maximum standard deviation,  rather than normalizing each component by its own standard deviation.  \CRCAr{Without normalization, the lower-order coefficients with larger magnitudes would dominate training, potentially causing instability and poor gradient updates. The mean subtracted corresponds to the time-averaged mean flow field from which the POD coefficients are derived, ensuring the data is centered before normalization.}

Figure \ref{auto_figure}a shows the relative error on the test data of the POD coefficients using standard and hybrid autoencoders when varying $d_h$ from 10 to 20. We have also added the corresponding values from a linear projection onto an equivalent number of POD coefficients. The POD projection onto the leading (complex) coefficients can be expressed as $\boldsymbol{a}=\boldsymbol{U}_r^T\boldsymbol{u}$.
\CRCAr{For each latent dimension $d_h$ two autoencoders are trained independently to reduce the effects of the inherently stochastic nature of neural network training, including random weight initialization and mini-batch sampling during optimization, which can lead to variability in performance across training runs. To mitigate this, multiple models are trained, and the one with the lowest validation error is selected to represent performance at that dimension.}
The same architectures are used for the standard and hybrid autoencoders (see table \ref{table_arch}). 

In Figure \ref{auto_figure}a, the nonlinear reduction leads to nearly one-order-of-magnitude decrease in the value of mean squared error (MSE) compared to its equivalent with POD for the same dimensionality.
Notably, a small reduction in error is observed beyond a threshold of $d_h>17$. 
\CRCAr{For POD, the relative error appears to plateau beyond this point, indicating convergence to a low-dimensional representation. In contrast, the autoencoders exhibit a more gradual reduction in error, which resembles a power-law decay rather than a distinct plateau. 
% Beyond this point, the relative error  appears to plateau with a minimum MSE reduction, suggesting that the autoencoders are capable of finding  a mapping from the full space to a low-dimensional representation.
This implies that, with dimensions as few as 17, the autoencoders  provide a good coordinate transformation from the full space  (e.g., as we will show in section \ref{modelling})}. 
% Next, we will compare the performance of autoencoders, but, we will  select $d_h=20$ to ensure that the extra degrees of freedom help to capture the complex dynamics of the system in the manifold coordinates. 
In all considered cases, the hybrid autoencoder consistently produces slightly better results in terms of MSE compared to the standard autoencoders. In Figure \ref{auto_figure}b, we compare the performance of both autoencoders by plotting the mean-squared POD coefficient amplitudes for the test data, denoted as  $\left \langle \left \| a_n \right \|^2 \right \rangle$, for the low-dimensional representation with $d_h=20$ (see figure \ref{auto_figure}b). The hybrid autoencoder exhibits limitations in capturing the amplitude of higher-order coefficients beyond $\boldsymbol{a}>30$, while the standard autoencoder struggles to accurately represent coefficients beyond the first two. This discrepancy arises from the fact that the \CRCAr{hybrid} autoencoder prioritizes the reconstruction of the leading $d_h$  POD coefficients.
Figure \ref{auto_figure}c illustrates the Reynolds stresses for both types of autoencoders  with $d_h=20$ for 5000 snapshots of the test data, showing that the hybrid autoencoder outperforms the standard autoencoder.
Finally, figure \ref{auto_figure}d displays field snapshots in the $z -\theta$ plane ($r = 0.496$) at random times showing qualitatively that hybrid autoencoders with $d_h=20$   can accurately reconstruct the data.
\CRCAr{This result agrees with the findings from
\citet{Kreilos}, who showed in plane Couette flow that high-dimensional systems can live in low-dimensional manifolds.}

Lastly, it is important to note that ideally the relative error value would plateau after determining the `right' manifold dimension yet it is not always feasible, primarily due to computational limitations introduced during the training process. Our goal is to identify the optimal dimension for constructing DManD models that faithfully capture both short-time tracking and long-time statistical characteristics of turbulent pipe flow. In the upcoming section, we will systematically build models with varying latent dimension sizes \CRCAr{(e.g., size of the low-dimensional space learned by the encoder)}. 

\CRCAr{ We compared models using 512 and 1024 POD modes and found that increasing to 1024 resulted in less than 0.02\% improvement in prediction accuracy (with $d_h=20$), despite a significant increase in training time. Since 512 modes already capture over 99.44\% of the total energy in comparison to 99.91\% for 1024 modes, we use 512 modes  for efficiency without loss of accuracy.}

\CRCAr{Although a deep autoencoder could, in theory, learn a low-dimensional representation directly from high-dimensional input (e.g., $\mathcal{O}(10^5)$), training dense networks at this scale is computationally expensive and prone to overfitting \citep{goodfellow2016deep}. To address this, we first apply POD to reduce the input dimensionality while preserving the dominant flow structures, enabling efficient and stable training. Alternatively, convolutional neural networks (CNNs) offer a scalable solution that can learn directly from high-dimensional data by exploiting local structure, potentially eliminating the need for POD. While we focus on POD-based preprocessing here, CNN-based models are the alternative \citep{fukami2019super}.}

%--------------------------------------
\subsection{Modelling in manifold coordinates \label{modelling}}
%--------------------------------------

Following the training of the autoencoders, a comprehensive exploration of the damping parameter $\gamma$ is conducted to prevent the dynamics from drifting away from the attractor during modelling. \CRCAr{ This is motivated by the concept of an inertial manifold: in dissipative systems such as the NSE, long-term dynamics are expected to collapse onto a lower-dimensional attracting set. In this spirit, the damping parameter acts as a regularizer that promotes stability in the latent dynamics by encouraging trajectories to remain close to the learned manifold.} To find the optimal value of $\gamma$, trials  are performed with $d_h=20$, varying $\gamma$ within the range $0.1 < \gamma < 0.5$. Empirical findings consistently point towards the fact that $\gamma=0.25$  yields superior outcomes, \CRCAr{with respect to the long-term dynamics of the system, as assessed by comparing the norm of the latent representation obtained from the autoencoder to that predicted by the NODE. We note that the significance of $\gamma$  has been rigorously investigated by \cite{linot_Couette}, in the context of  MFU plane Couette flow, who showed that without the damping term almost all models become unstable for longer runs.}
The training objective of the NODE is focused on predicting one time unit ahead  ($\tau = 1$), as described by equation \ref{eq:Node2}.
As a preprocessing step for training the NODE, \CRCAr{we subtract the mean of the autoencoder's  latent representations to center the data}. This centralization ensures that the linear damping effectively guides trajectories towards the origin.

Unless otherwise specified, the results presented showcase the top-performing model at each dimension, with the lowest relative error averaged across all considered statistics. It is crucial to note that, for all DNS versus DManD model comparisons at each $d_h$, identical initial conditions are applied in the models. From the perspective of the DManD models, this involves encoding the initial condition from the DNS, and subsequently evolving it forward \CRCAr{in} time in the latent space to generate a time series of $\boldsymbol{h}$.  This time series is then decoded to the full state space for comparative analysis.

\subsubsection{Short-time tracking}

\begin{figure}
\begin{center}
\begin{tabular}{cc}
\includegraphics[width=0.4\linewidth]{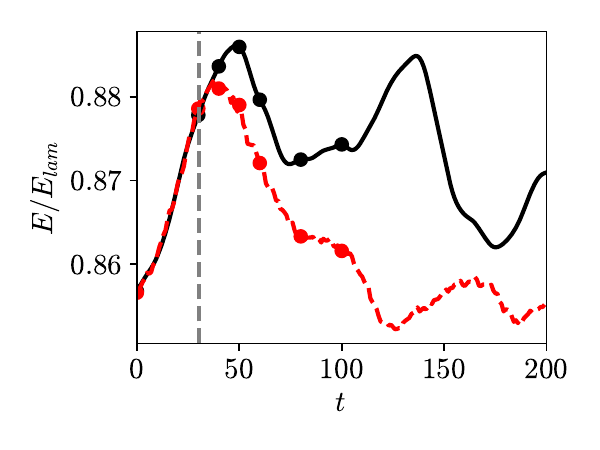}&
\includegraphics[width=0.4\linewidth]{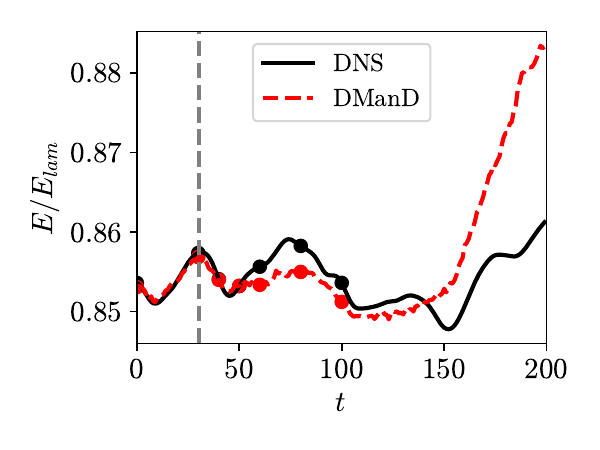}\\
 (a) & (b)\\
\end{tabular}
\begin{tabular}{c}
\includegraphics[width=\linewidth]{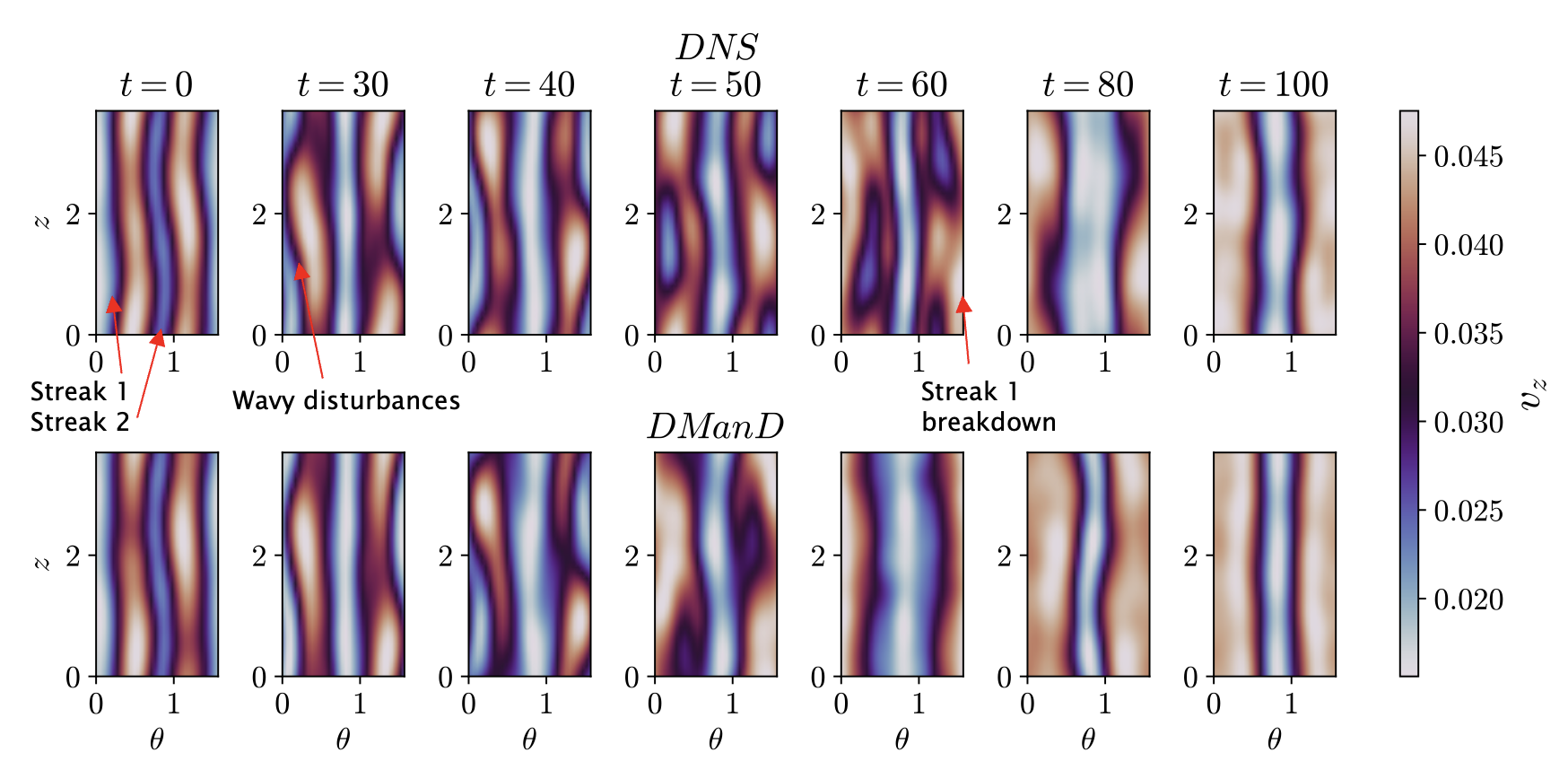}\\
(c)\\
% \includegraphics[width=0.55\linewidth]{Fig/DManD/E_406_norm.pdf}\\
% % (a)\\
\includegraphics[width=\linewidth]{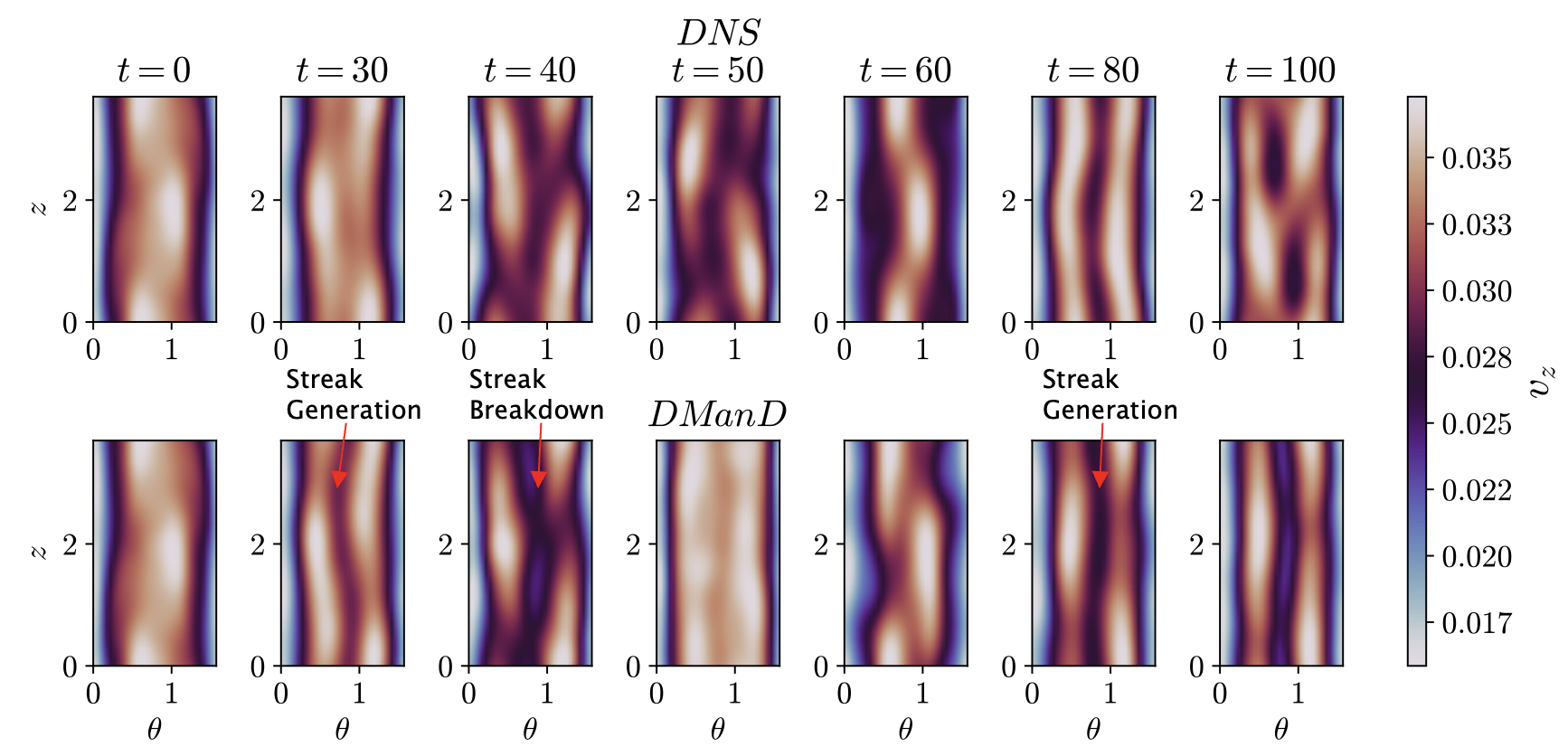}\\
(d)
\end{tabular}
\end{center}
\caption{\label{ssp_figure}  Normalised kinetic energy of the system for the DNS and DManD model with $d_h = 20$ up to $t=200$ \CRCAr{shown for two random initial conditions, corresponding to panels (a) and (b), respectively)}.  \CRCAr{Panels (c) and (d) represent } two-dimensional representation of the dynamics in the $z-\theta$ plane ($r = 0.496$) with $u_z$ for the DNS and DManD model \CRCAr{for IC corresponding to panels (a) and (b), respectively}. The vertical dashed line marks one Lyapunov time.
We refer the reader to the supplemental materials  to view a video of the trajectory corresponding to panel (a).} 
\end{figure}

In this section, we evaluate the performance of the DManD models in reconstructing short-time trajectories. Figure \ref{ssp_figure}a shows the time evolution of the kinetic energy of the system for the DNS and  DManD model with $d_h=20$. Here, the kinetic energy of the system is given by the $L^2$ inner product

\begin{equation}
    E=\frac{1}{\mathcal{V}}\int_{0}^{2\pi/\alpha}\int_{0}^{2\pi/m_p}\int_{0}^{\CRCAr{R}} \frac{1}{2}\boldsymbol{v}\cdot \boldsymbol{v}~ r ~ dr d\theta dz
\end{equation}
where $\mathcal{V}$ corresponds to \CRCAr{the volume of the} cylindrical flow domain.  The results displayed in figure \ref{ssp_figure}a  are normalized by the kinetic energy of the laminar state.
We have selected $d_h = 20$ as it represents the minimum dimension that yields superior results for both short-term and long-time measures, as will be demonstrated in this and the subsequent sections. We note that  $d_h = 20$ may not necessarily correspond to the exact dimension of the manifold, but this corresponds to the smallest dimension that can faithfully capture the nonlinear dynamics of the turbulent pipe flow \CRCAr{in the present modelling framework}. We reiterate that we have started from an initial state dimension of $\mathcal{O}(10^5)$, and developed a dara-driven model of $\mathcal{O}(10)$ without substantial loos of accuracy. Figure \ref{ssp_figure}a,b show that the DManD model can generate predictions that capture the true dynamics of the system up to $t\sim 50$, \textcolor{black}{ which corresponds to slightly more than one Lyapunov time ($t_L= 30.43$, this value is calculated later from the DNS)}. To provide a more qualitative representation of the model dynamics, figure \ref{ssp_figure}\CRCAr{c,d} display a two-dimensional representation of the dynamics in the $z-\theta$ plane ($r = 0.496$) with $v_z$ for the DNS and DManD model \CRCAr{for IC corresponding to panels \ref{ssp_figure}a and \ref{ssp_figure}b, respectively}. We observe a good agreement between the model and the DNS. We refer the reader to the supplementary materials to view a video from the trajectory for \CRCAr{figure \ref{ssp_figure}a}.

\begin{figure}
\begin{center}
\begin{tabular}{ccccc}
$t=0$ & $t=20$ & $t=40$ & $t=100$ & $t=180$\\
\includegraphics[width=0.18\linewidth]{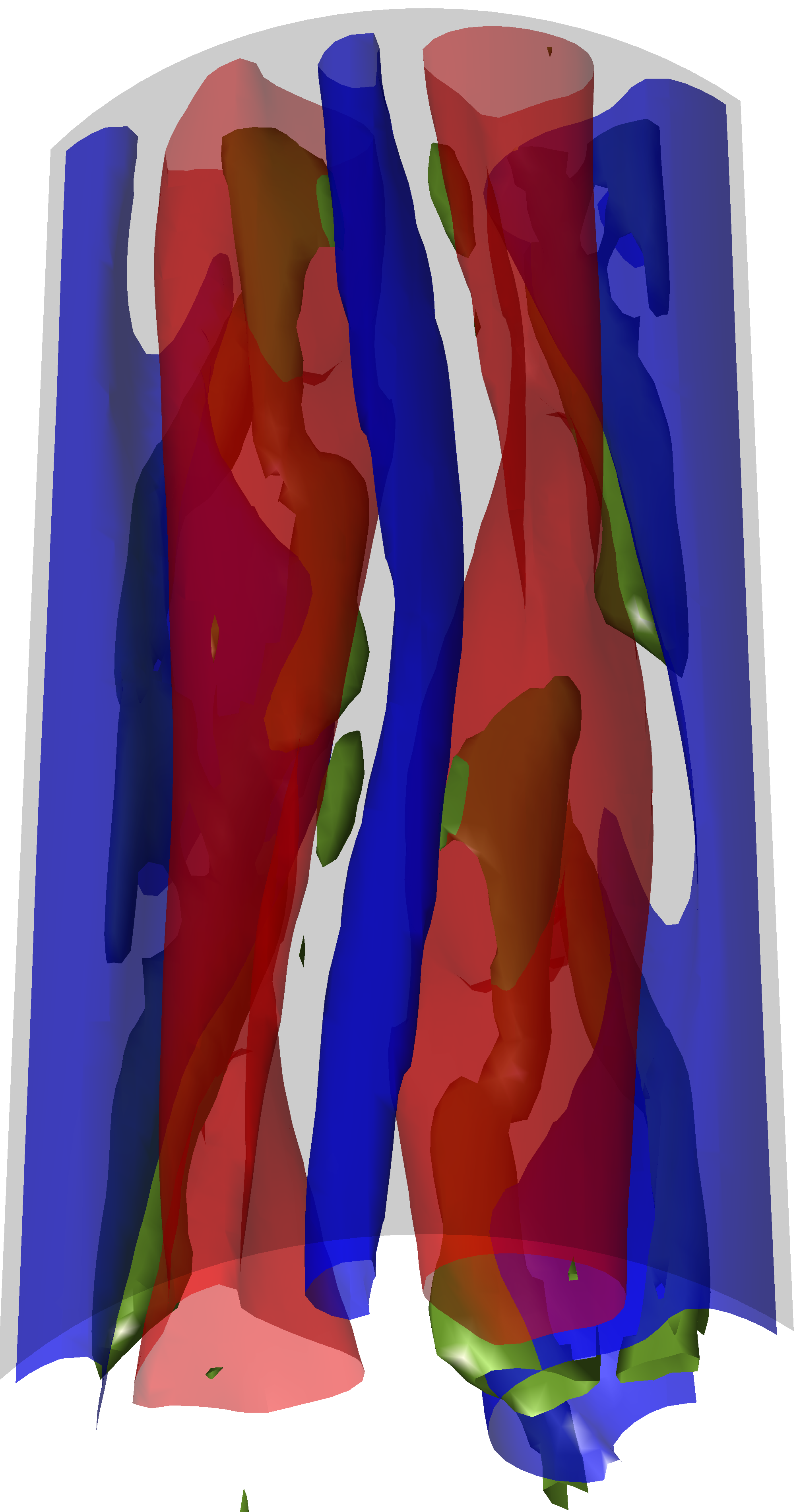}&
\includegraphics[width=0.18\linewidth]{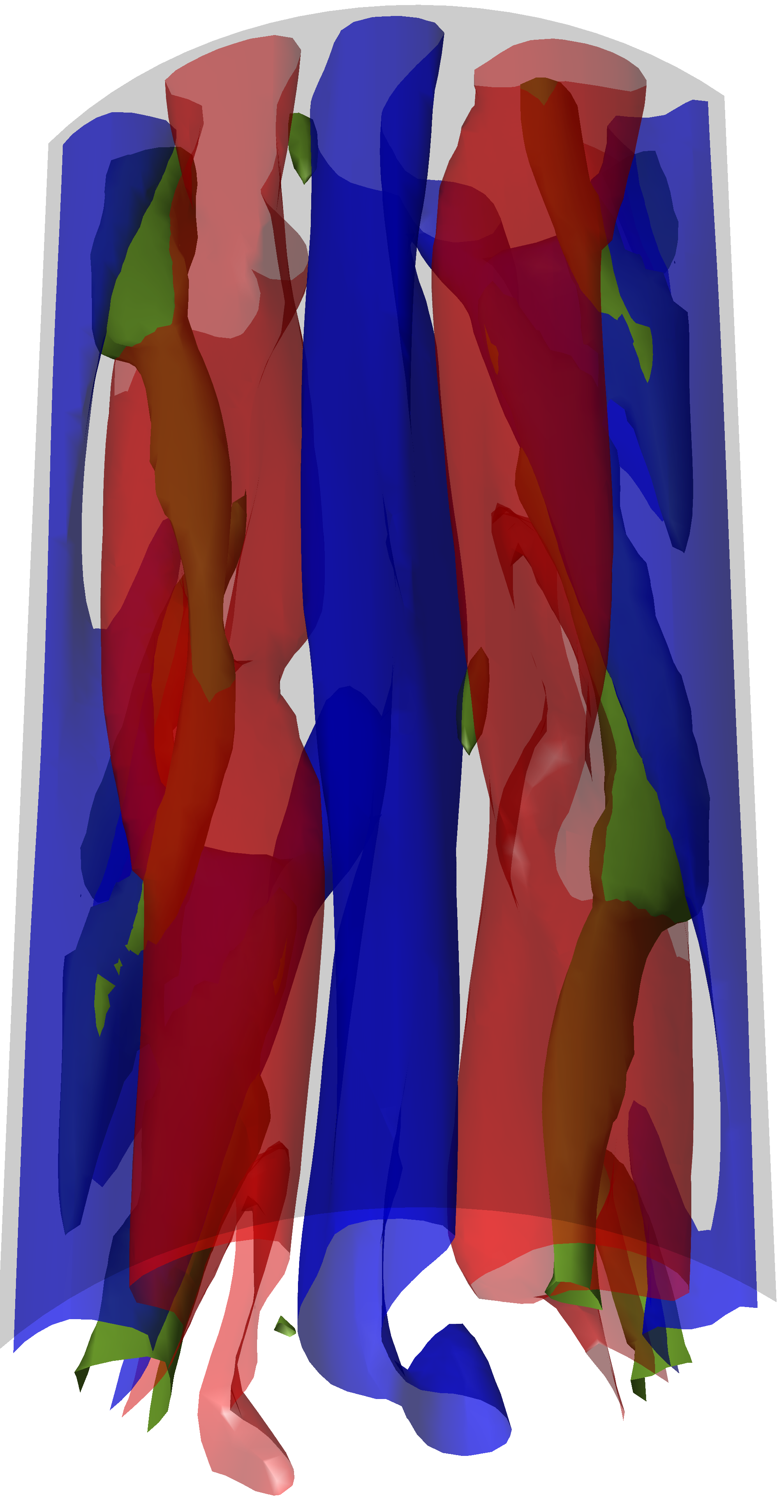}&
\includegraphics[width=0.18\linewidth]{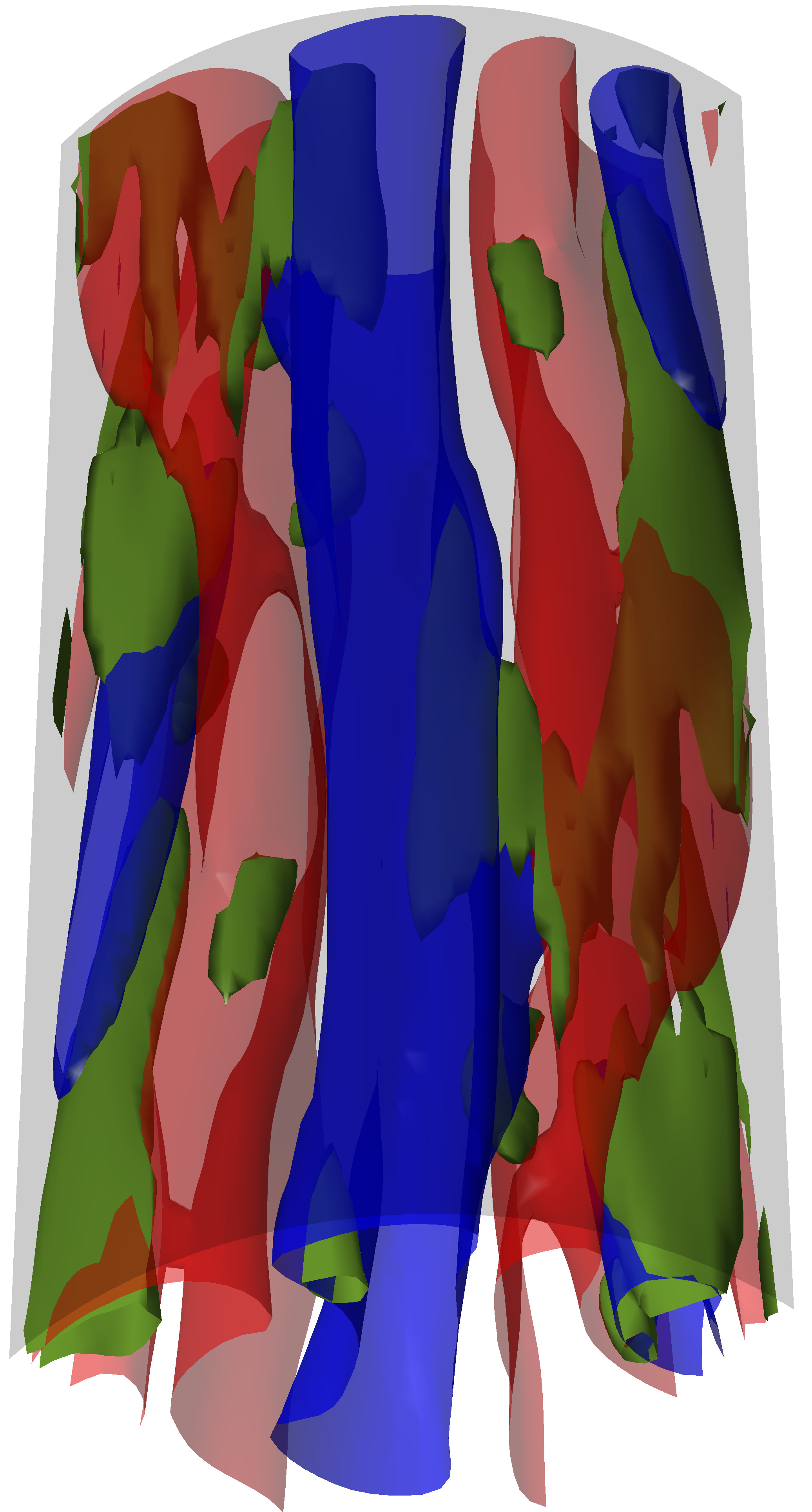}&
\includegraphics[width=0.18\linewidth]{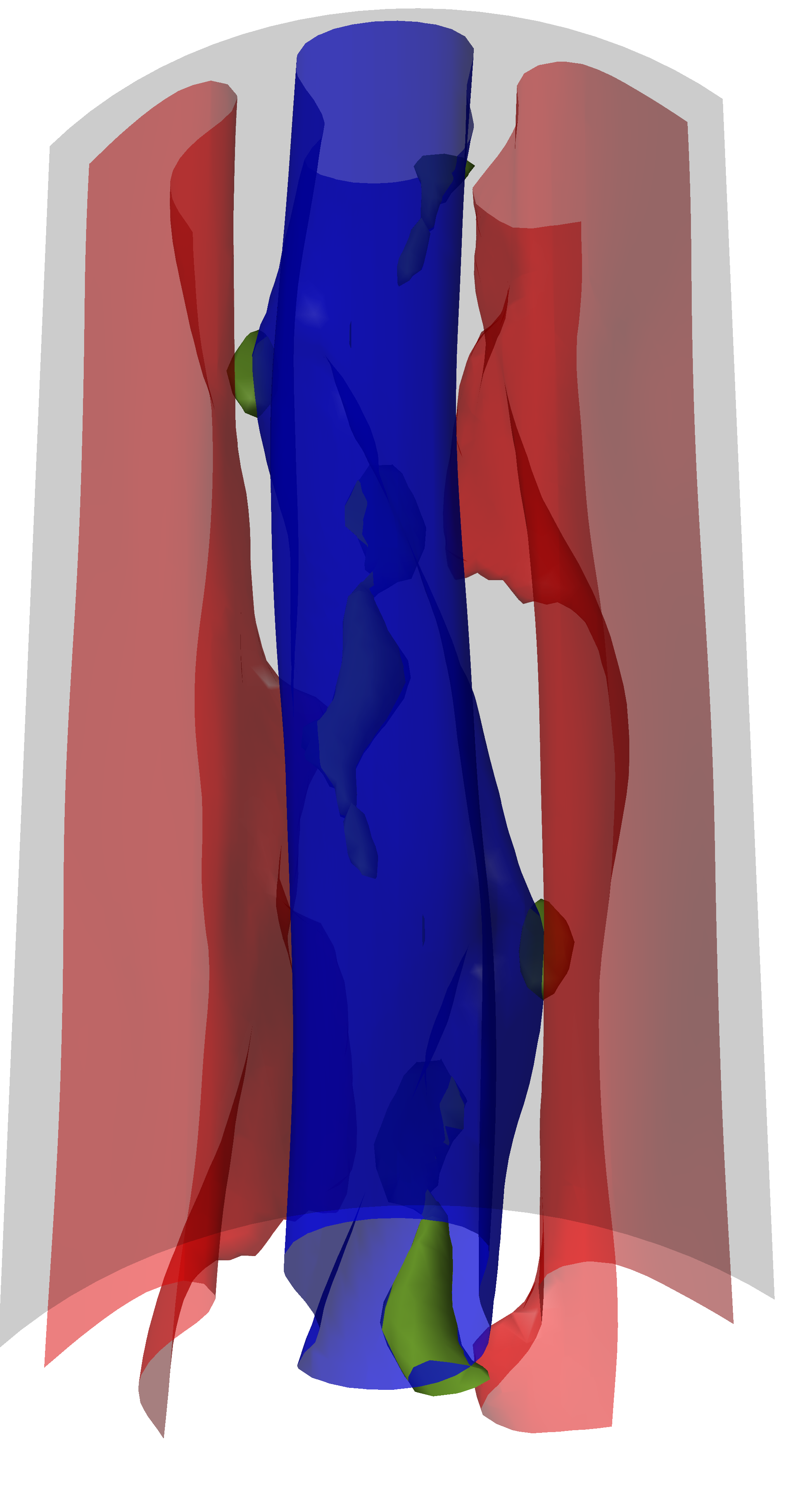}&
\includegraphics[width=0.18\linewidth]{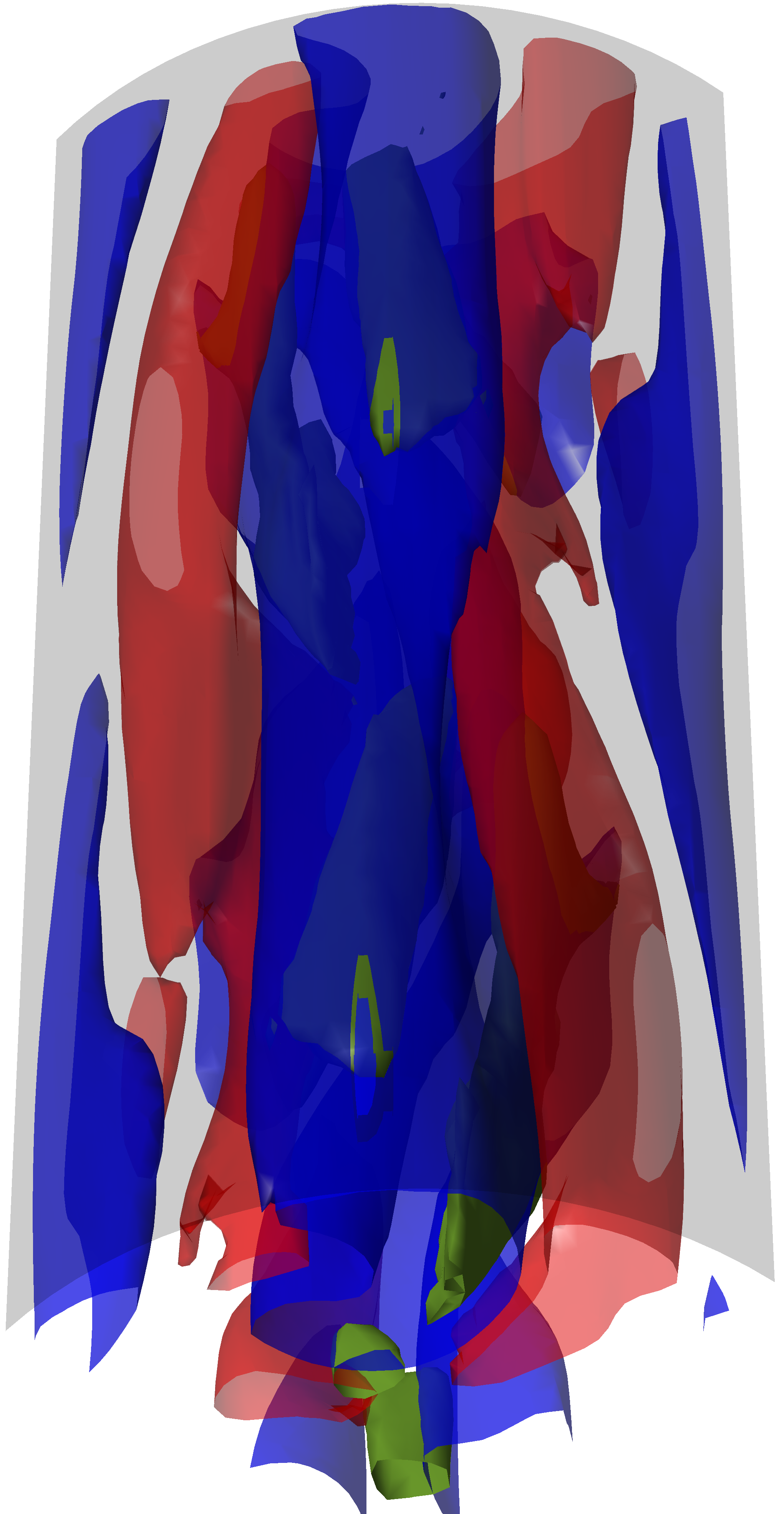}\\
\parbox[c]{0.18\linewidth}{\centering \scriptsize Well-defined \\ streaks} &
\parbox[c]{0.18\linewidth}{\centering \scriptsize Wavy instabilities/ \\ begin streak breakdown} &
\parbox[c]{0.18\linewidth}{\centering \scriptsize Emergence \\ of vortices} &
\parbox[c]{0.18\linewidth}{\centering \scriptsize Regeneration \\ of streaks} &
\parbox[c]{0.18\linewidth}{\centering \scriptsize Repeat  wave formation/ \\ process continues}
% &&(a)&&\\
% $t=0$ & $t=20$ & $t=30$ & $t=40$ & $t=100$\\
% \includegraphics[width=0.15\linewidth]{Revision/406_0_SSP.png}&
% \includegraphics[width=0.15\linewidth]{Revision/406_20_SSP.png}&
% \includegraphics[width=0.15\linewidth]{Revision/406_30_SSP.png}&
% \includegraphics[width=0.15\linewidth]{Revision/406_40_SSP.png}&
% \includegraphics[width=0.15\linewidth]{Revision/406_100_SSP.png}\\
% &&(b)&&\\
 % (a) & (b) & (c) & (d)\\
\end{tabular}
\end{center}
\caption{\label{SSP} \textcolor{black}{ Self-sustaining process in DManD model corresponding to the initial condition shown in Figure 4a. Isosurfaces of the streamwise velocity fluctuations are displayed for $ u_z' = 0.05 $ (blue, representing fast streaks) and $ u_z' = -0.05 $ (red, representing low-speed streaks). Additionally, each snapshot includes isosurfaces of the $\lambda_2$ criterion with a threshold of $ \lambda_2 = 0.1 $, shown in green, to highlight vortical structures.
% In the bottom panels, the thresholds are lowered to $u_z' = 0.01$ and $u_z' = -0.01$, highlighting weaker velocity fluctuations. 
Only a quarter of the domain is shown. \label{SSP_iso} }}
\end{figure}

\textcolor{black}{The initial conditions for Figure \ref{ssp_figure} are chosen to demonstrate that DManD qualitatively captures the dynamics observed in the true data. These two initial conditions are representative of the Self-Sustaining Process (SSP), in which streamwise rolls, streaks, and wavelike disturbances mutually reinforce one another, thereby counteracting viscous decay. This behavior is further supported by the results shown in figure~\ref{SSP_iso}. } 
% The initial conditions for figure \ref{ssp_figure} are selected because  they highlight the Self-Sustaining Process (SSP) where streamwise rolls, streaks, and wavelike disturbances mutually sustain each other, counteracting viscous decay. 
In \CRCAr{panel \ref{ssp_figure}c} the true system (top panels), streaks (velocity \CRCAr{fluctuations} below the mean) near the wall \CRCAr{(see blue color at the edges of the panel} at $t=0$) are affected by azimuthal wavy disturbances \CRCAr{(at $t=30-40$)}, leading to their breakdown \CRCAr{(at $t=50$)} and the formation of rolls (after $t=80$). DManD qualitatively captures these dynamics (see the corresponding bottom panels) with a slightly accelerated breakdown of the rolls into streaks. \CRCAr{In panel~\ref{ssp_figure}d, we track a second initial condition that also  highlights some parts of the SSP dynamics. At $t = 30$,  streaks are observed near the center of the domain. These streaks undergo a breakdown and temporarily vanish by $t = 50$, before gradually reemerging by $t = 80$. This cyclical pattern observed in panel a and b, decay followed by regeneration, reflects the characteristic SSP interplay between streamwise rolls, streaks, and wavelike instabilities. The DManD reconstruction (bottom rows) qualitatively reproduces this sequence of events, capturing the key  transitions present in the DNS (top rows).
}
% near the wall are influenced by azimuthal wavy disturbances (see $t=30$). These disturbances lead to the breakdown of the streaks (see $t=50$) and the formation of rolls (after $t=70$). The rolls, in turn, induce the re-emergence of the streaks (see $t=350$).

% \begin{figure}[h!]
% \centering
% % Row of images
% \begin{tabular}{ccccc}
% $t=0$ & $t=20$ & $t=40$ & $t=100$ & $t=180$\\
% \includegraphics[width=0.18\linewidth]{Revision/271_ssp_0.png} &
% \includegraphics[width=0.18\linewidth]{Revision/271_ssp_20.png} &
% \includegraphics[width=0.18\linewidth]{Revision/271_ssp_40.png} &
% \includegraphics[width=0.18\linewidth]{Revision/271_ssp_100.png} &
% \includegraphics[width=0.18\linewidth]{Revision/271_ssp_180.png} \\
% \small Well-defined streaks &
% \small Wavy instabilities / onset of breakdown &
% \small Emergence of vortices &
% \small Regeneration of streaks &
% \small Wave formation continues
% \end{tabular}
% \caption{Temporal evolution of the flow.}
% \label{fig:ssp_panels}
% \end{figure}

\textcolor{black}{Figure \ref{SSP_iso} illustrates the dynamics of the Self-Sustaining Process (SSP) observed in the DManD model using isosurfaces for the streaks and vortical structures. The blue and red isosurfaces represent high ($u_z' = 0.05$)- and low ($u_z' = -0.05$)-speed streaks, respectively, while the green isosurfaces correspond to regions of high vorticity, identified using the $\lambda_2$ criterion. At $t=0$, the flow exhibits well-defined streamwise streaks. Due to the imposed periodic boundary conditions, only one quarter of the domain is shown, effectively capturing a pair of streaks. By $t=20$, the low-speed streaks near the lateral walls begin to exhibit sinuous (wavy) instabilities, which amplify and lead to streak breakdown. This breakdown gives rise to streamwise vortices, clearly visible at 
$t=40$ through the emergence of green $\lambda_2$ criterion structures. As the cycle progresses, these vortical structures redistribute momentum and energy, leading to the regeneration of coherent streaks by 
$t=100$. At $t=180$, a new breakdown event is underway, completing one full SSP cycle. This sequence captures the essential feedback loop of streak formation, instability, breakdown, and regeneration that underpins sustained turbulence in wall-bounded flows.}

\begin{figure}
\begin{center}
\begin{tabular}{c}
\includegraphics[width=0.8\linewidth]{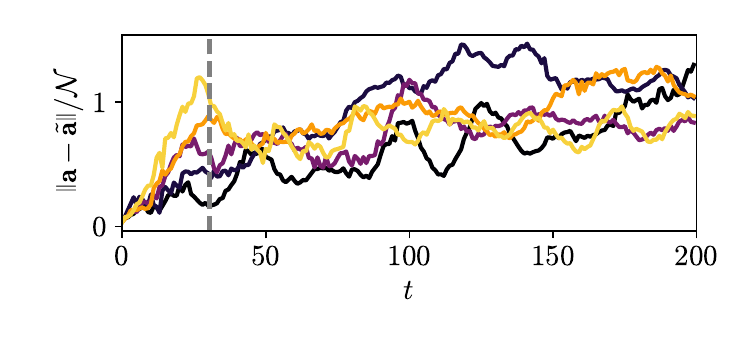}\\
\end{tabular}
\end{center}
\caption{\label{IC_short_time} 
Normalised short-time-tracking error for five arbitrary initial conditions using DManD with $d_h=20$ up to $t=200$. The vertical dashed line marks one Lyapunov time. \CRCAr{Two of the lines corresponds to the initial conditions shown in figure 4.} }
\end{figure}

While figure \ref{ssp_figure} only shows the trajectory for \CRCAr{two} selected initial conditions,  figure \ref{IC_short_time} represents the tracking error for five random initial trajectories at $d_h=20$. Here we plot $\left \| \boldsymbol{a}(t)- \tilde{\boldsymbol{a}}(t) \right \|_2^2/\mathcal{N} $, where $\mathcal{N}$ denotes the error of true solutions at random times  $t_i$ and $t_j$, i.e., $\mathcal{N}=\left \langle   \left \|  \boldsymbol{a}(t_i) - \boldsymbol{a}(t_j) \right \| \right \rangle$.
To enhance computational efficiency, we opted for comparisons in POD coefficients space rather than reconstructing full velocity fields. This decision was motivated by the computationally intensive nature (i.e., memory usage) of reconstructing fields from POD coefficients. Moreover, the POD coefficients enable the capture of the $99.83\%$  of the energy within the system (with 512 POD modes).
Figure \ref{IC_short_time} shows that for certain initial conditions, the error remains relatively low until $t=50$, after which it increases before stabilizing at unity at longer times.
\CRCAr{ This behavior is expected for chaotic systems, where small initial errors grow exponentially and eventually lead to complete decorrelation between predicted and true trajectories. Once this occurs, the normalized error saturates at its maximum possible value, indicating a loss of predictive capability.}
% in line with expectations. 
\CRCAr{
We note the recent work by \citet{Vela_Avila_2024}, who demonstrated that in Kolmogorov flow, the short-term predictability limit for a given uncertainty in initial conditions depends strongly on the location of the initial condition within the attractor. This supports our observation in Figure \ref{IC_short_time} that predictive accuracy varies across different initial conditions due to the intrinsic structure of the chaotic attractor.}

\begin{figure}
\begin{center}
\begin{tabular}{cc}
\includegraphics[width=0.5\linewidth]{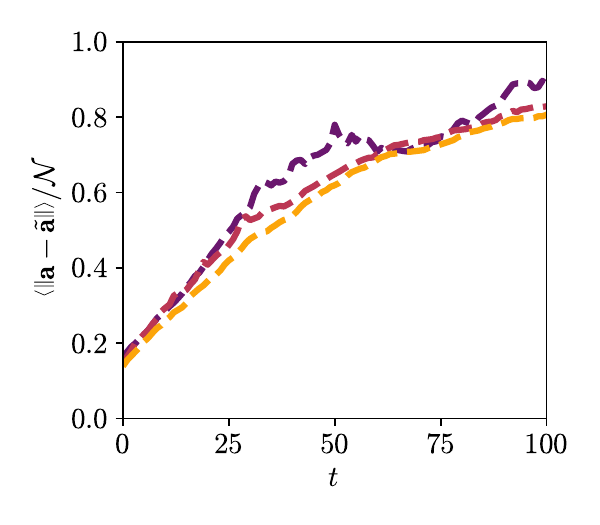}&
\includegraphics[width=0.5\linewidth]{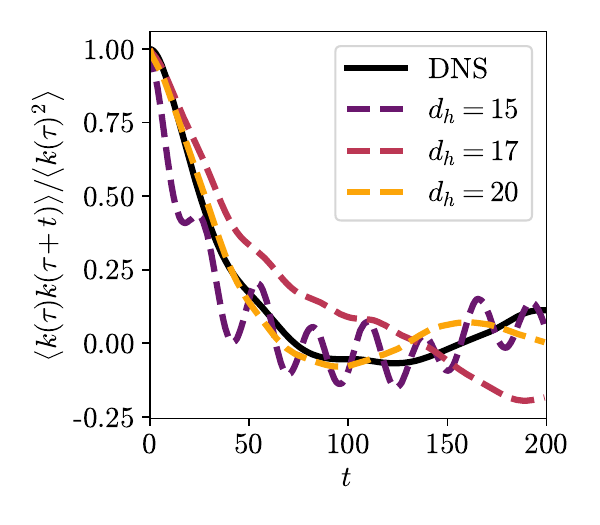}\\
(a) & (b)\\
\end{tabular}
\end{center}
\caption{\label{short_time_auto} 
Short-time tracking performance: (a) Ensemble-average of 500 random initial conditions as a function of $d_h$. (b) Temporal autocorrelation of fluctuations in the kinetic energy as a function of $d_h$. The black solid line represents the temporal autocorrelation calculated in the DNS. For representation purposes, we only show results for $d_h=[15,17,20]$. }
\end{figure}

Next, we conduct a parametric study on the normalized ensemble-averaged tracking error for DManD models by varying the dimension of the latent space (see figure \ref{short_time_auto}a). We used 500 random initial conditions evolved over 100 time units. Notably, we observe that when the dimensionality surpasses 17 ($d_h > 17$), the tracking errors converge, indicating that a dimension of at least 17 is required for a better field reconstruction.

To understand the short-time tracking and the correlation of the models with different initial conditions, we  plot the autocorrelation of fluctuations of the kinetic energy.  We define its fluctuating part as $k(t)=E(t)-\langle E \rangle$. In figure \ref{short_time_auto}, we plot the temporal autocorrelation of $k$ with respect  to its corresponding initial condition for 6000 random initial conditions evolved up to $t=200$. It is not until $d_h=20$, that the predicted autocorrelation matches reasonably well up to $t\sim 120$ with respect to the true data.

\subsubsection{Long-time statistics}

\begin{figure}
\begin{center}
\begin{tabular}{c}
\includegraphics[width=0.5\linewidth]{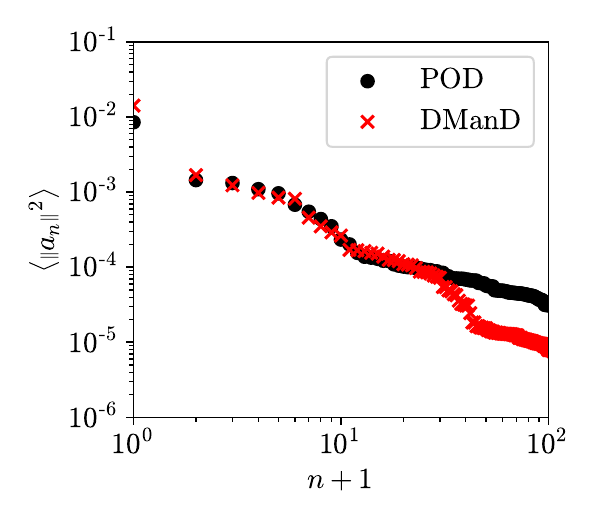}
\end{tabular}
\end{center}
\caption{\label{amp_pod_dh20} 
 Comparison of $\left \langle \left \| a_n \right \|^2 \right \rangle$ for the  DNS and DManD at $d_h = 20$ for the same inital condition evolved 3000 time units.} 
\end{figure}

This section is dedicated to presenting the long-time statistics predictions for the DManD models. To evaluate the long-time performance of the DManD models, we first use the $\left \langle \left | a_n \right |^2 \right \rangle$ metric, plotting it for a long trajectory (up to 3000 time units) in both the DNS solution and the predicted trajectories from DManD at $d_h=20$ (see figure \ref{amp_pod_dh20}). We observe that DManD shows good agreement with the true solution up to the first $30$ leading POD coefficients of the true solution. This suggests that DManD effectively captures  the attractor structure, and subsequently the temporal prediction of the nonlinear dynamics of the system over an extended time span. To benchmark the performance of our framework, we draw parallels with classical methods for reduced order models proposed by \citet{Gibson} for Couette flow (i.e., POD-Galerkin).  \citet{Gibson} required between 512 ($\sim$ 1000 degrees of freedom) and 1024 modes ($\sim$ 2000 degrees of freedom) to achieve a reliable prediction of the leading 30 POD coefficients for plane Couette flow.

\begin{figure}
\begin{center}
\begin{tabular}{c}
\includegraphics[width=\linewidth]{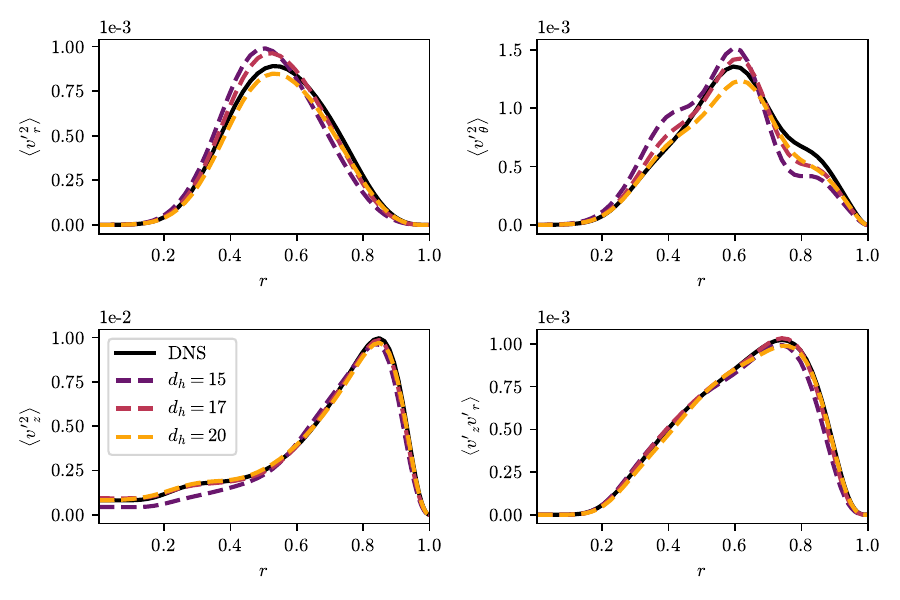}\\
\end{tabular}
\end{center}
\caption{\label{Reynolds_Stresses} 
 Long-time statistics: Components of the Reynolds stresses with increasing dimension for DManD models at various dimensions. The black solid lines represent the Reynolds stresses calculated in the DNS. For clarity, we only show results for $d_h=[15,17,20]$.} 
\end{figure}

\begin{figure}
\begin{center}
\begin{tabular}{c}
\includegraphics[width=\linewidth]{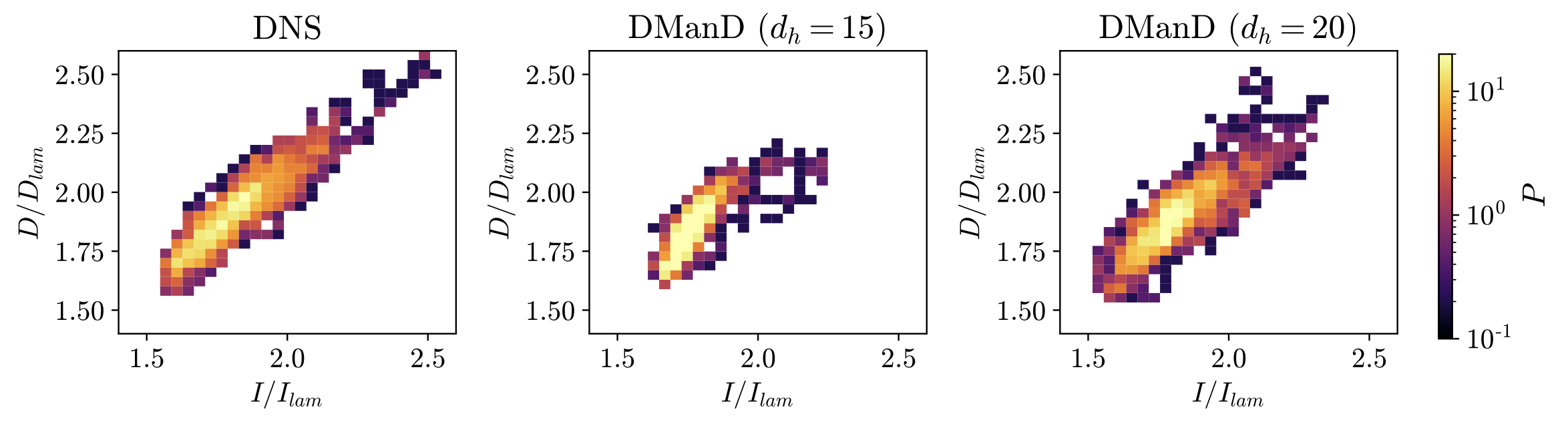}\\
(a)\\
\includegraphics[width=0.45\linewidth]{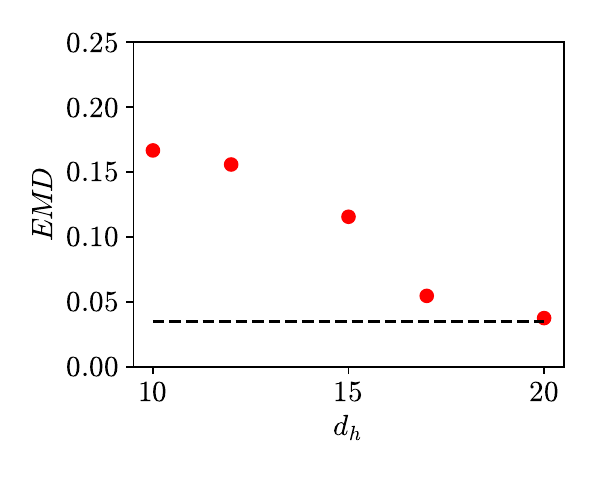}\\
\textcolor{black}{(b)}
\end{tabular}
\end{center}
\caption{\label{EID} Energy balance: (a) Joint PDFs of the dissipation ($D$) and power input ($I$) for the true system, and the DManD models at $d_h = 15$ and $d_h = 20$, corresponding to columns one to three, respectively. (b) Earth movers distance (EMD) between the PDF from the DNS and the PDF predicted by the DManD model at various dimensions $d_h$. The dashed line represents the error between two PDFs generated from DNS trajectories of the same length but with different initial conditions. } 
\end{figure}

We turn our attention to the predictions of various models concerning the Reynolds stresses, as illustrated in figure \ref{Reynolds_Stresses}. The streamwise velocity component  $\left \langle u^2_z \right \rangle$ is the most important component with a peak near the wall, being one order of magnitude bigger than the other components. We observe that our  DManD models perform well in predicting the Reynolds stresses of the system beyond $d_h > 17$. At $d_h = 20$, in particular, DManD exhibits exceptional performance by closely matching three out of the four displayed components. However, minor discrepancies are observed in the component corresponding to $\left \langle u^2_\theta \right \rangle$.

Next, we assess how well the DManD models are capable of reconstructing the energy transfer rates at long times by examining the joint probability density functions of power input ($I$) and dissipation ($D$). The power input required to maintain constant  mass flux and dissipation due to viscosity  are defined as
\begin{equation}
I=\frac{1}{\mathcal{A}}\int_{0}^{2\pi/\alpha}\int_{0}^{2\pi/m_p}  \left | \frac{\partial v_z}{\partial r}\right |_{r=R} ~ d\theta dz,
\end{equation}
\begin{equation}
D=\frac{1}{\mathcal{V}}\int_{0}^{2\pi/\alpha}\int_{0}^{2\pi/m_p}\int_{0}^{\CRCAr{R}} | \bigtriangledown \times \boldsymbol{v}|^2~ r ~ dr d\theta dz.
\end{equation}
here $\mathcal{V}$ and $\mathcal{A}$ stand for the volume \CRCAr{of the cylindrical flow domain} and area of the pipe, respectively. In the results shown in this paper,  the energy input and dissipation values are normalized with respect to their laminar values. An energy balance can be derived from the inner product $\langle \boldsymbol{v}, \partial \boldsymbol{v}/\partial t \rangle$ \citep{waleffe_2001}. 
Then, the energy input rate, the dissipation rate, and the kinetic energy  $ \dot{E}  = (I -D)/Re$, which must average to zero over long times. It is imperative to ensure that this quantity averages to zero over extended periods, signifying equilibrium. This assessment is important for assessing the accuracy of DManD models in maintaining energy balance. 
\CRCAr{Specifically, we define the energy balance deviation  as
$\mathrm{EB} = \left\langle |I(t) - D(t)| \right\rangle/Re$,
where \(I(t)\) and \(D(t)\) represent the cumulative energy input and dissipation up to time \(t\). Over a trajectory of 5000 time units, the average deviation remains below 1\%, demonstrating that the model faithfully captures the essential energy transfer and dissipation mechanisms without explicitly enforcing energy conservation.}

Figure \ref{EID} displays the joint PDFs of the normalised  $D$ and  $I$ from the DNS, and  DManD models with $d_h=15$ and $d_h=20$, generated from a long time trajectory evolved up to 3000 time units (with the same initial condition). For  $d_h=15$, the model fails to capture the intrinsic nonlinear dynamics of the system, but for  $d_h=20$, the model accurately captures the core region of these projections, including the excursions occurring at high dissipation rates that are also present in the DNS results, \CRCAr{indicating that high-dissipation bursts are preserved in the learned manifold when $ d_h $ increases. This suggests that the model is capable of encoding rare events that are observed in the DNS.} Overall, we can conclude that DManD can effectively predict accurately the long-time statistics of this complex system in the coordinates of the low-dimensional representation.

To further quantify the divergence between the PDFs from the DNS and DManD, we calculate the earth movers distance (EMD) as a function of the dimension of the low-dimensional model.
The EMD  measures the distance between two PDFs by framing the true PDF as the `supplies' and the DManD model PDF as the `demands' \CRCAr{\citep{emd_reference}. 
 We use the EMD as a robust and interpretable metric of similarity between probability distributions. Unlike Kullback-Leibler divergence, EMD is a true distance that captures both the magnitude and spatial displacement of probability mass- features that are especially relevant in turbulent flows, where dissipation can exhibit heavy tails or abrupt shifts across regimes.}
EMD seeks to minimize the effort required to transport the supplies to meet the demands, essentially solving a transportation problem. We find the flow $f_{ij}$ that minimises $\sum_{i=1}^{m} \sum_{j=1}^{n} f_{ij} d_{ij}$ subject to the constraints

\begin{equation}
 f_{ij} \geq 0, ~~1\leq i \leq m, ~~ 1\leq j \leq n, 
\end{equation}
\begin{equation}
   \sum_{j=1}^{n}f_{ij}=p_i ~~1\leq i \leq m, 
\end{equation}
\begin{equation}
   \sum_{i=1}^{m}f_{ij}=q_i ~~1\leq j \leq n.
\end{equation}

Here, $p_i$ represents the probability density at the $i$th bin in the model PDF, and $q_j$ is the probability density at the $j$th bin in the DNS PDF, where both PDFs are discretized into $n$ and $m$ bins (in this scenario, $n = m$). Additionally, $d_{ij}$ denotes the cost associated with moving between bins, with the $L_2$ distance between bins $i$ and $j$ serving as the measure (where $d_{ij} = 0$ for $i = j$). After solving the minimization problem to determine the optimal flow $f^*_{ij}$, the EMD is calculated  as 

\begin{equation}
    EMD=\frac{\sum_{i=1}^{m}\sum_{j=1}^{n}f^*_{ij}d_{ij}}{\sum_{i=1}^{m}\sum_{j=1}^{n}f^*_{ij}}
\end{equation}

Figure \ref{EID}b shows the EMD values depending on the latent dimensions, with all DManD models starting from the same initial condition and evolved up to 3000 time units. Additionally, we include the error when comparing two PDFs generated from the DNS  with different conditions (dashed line). Figure \ref{EID}b demonstrates that the addition of latent dimensions results in an enhancement of the EMD value. Specifically, for $d_h=20$, the DManD model reaches a level of comparability to the DNS. This observation, combined with short-time tracking, supports the assertion that only 20 degrees of freedom are necessary to create low-dimensional models that faithfully capture both the \CRCAr{short-time tracking and long-time statistics} of the nonlinear turbulent dynamics of MFU pipe flow at $Re=2500$. It is important to note that we are not asserting a specific dimension for the manifold, but rather identifying the minimum dimension needed to produce accurate models. Similar results have been observed for plane MFU Couette flow \citep{linot_Couette,koopman_Couette} and Kolmogorov flow \citep{carlos_prf}.

\begin{figure}
\begin{center}
\begin{tabular}{cc}
\includegraphics[width=0.5\linewidth]{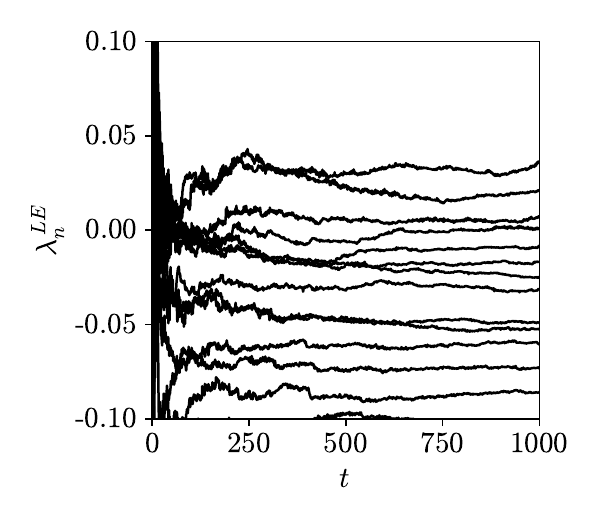}&
\includegraphics[width=0.5\linewidth]{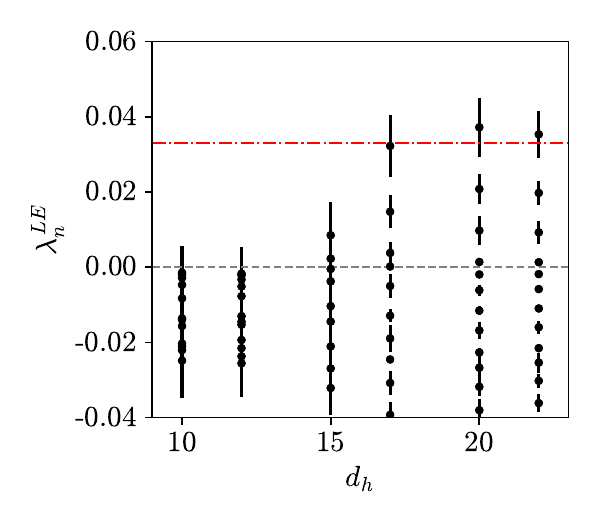}
\\
(a) & \textcolor{black}{(b)}\\
\end{tabular}
\begin{tabular}{c}
\includegraphics[width=0.5\linewidth]{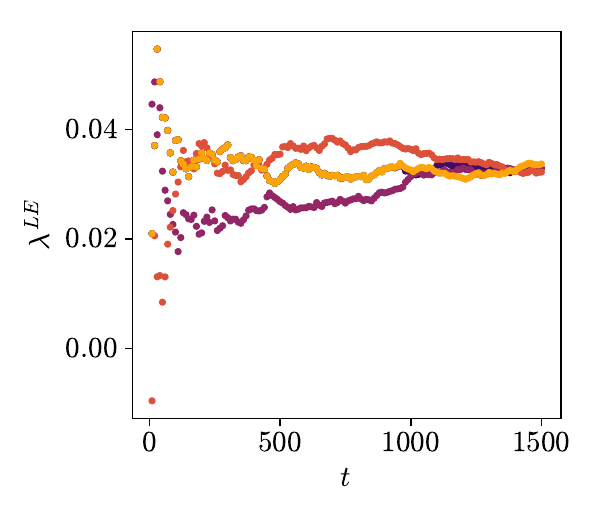}
\\
(c)\\
\end{tabular}

\end{center}
\caption{\label{Lyapunov}
(a)  Lyapunov exponents for the DManD models from a single trial with $d_h=20$. (b)   Lyapunov exponents for the DManD models at various dimensions with error bars representing the results from five different trials. The grey dashed line identifies $\lambda_n^{LE}=0$, \CRCAr{while the red dashed line represents the leading Lyapunov exponent of the DNS}. \CRCAr{(c) Leading Lyapunov exponent for the DNS for four different initial conditions.}} 
\end{figure}

Finally, we examine the leading  Lyapunov exponents of the DManD models depending on $d_h$.  The methods used are those described in \citet{Sandri}, with publicly available code from \citet{Rozbeda}. Simulations were run for over 1000 time units to ensure \CRCAr{convergence} of the Lyapunov exponents.  Figure \ref{Lyapunov}a shows a representative spectrum of Lyapunov exponents $\lambda_n^{LE}$  as a function of time, obtained from the DManD model with $d_h=20$ for a single initial condition. We effectively see three positive exponents, we also observe two exponents near zero due to the spatial translational symmetries in $\theta$ and $z$. Figure \ref{Lyapunov}b displays the exponents, averaged over five different initial conditions, as we vary the dimension of the DManD model.
Increasing the model dimension leads to enhancements in the estimation of these Lyapunov exponents. At low dimensions ($d_h \leq 12$), we do not observe any positive  Lyapunov exponents (i.e., models land in a fixed point). 
\CRCAr{
We also report the Lyapunov spectrum for $d_h = 22$ to demonstrate that the spectrum converges with increasing latent dimension $d_h$.
Furthermore, the leading Lyapunov exponent  computed from the DManD model closely matches that of the DNS (shown below), indicating that the dominant chaotic dynamics are well captured.}
% We note that we do not observe \CRCAr{convergence} on the Lyapunov exponents as the dimension increases.

\CRCAr{To enable comparison with DNS, we calculate the leading Lyapunov exponent (LLE) using the DNS solver. To do this calculation, we numerically evolve two nearby trajectories  in Openpipeflow, and the divergence rate of their separation over time is calculated. 
As the trajectories separate, the difference between them (or the perturbation) can grow or shrink significantly. To prevent numerical errors and ensure consistent tracking, the separation is rescaled every 10 time units to a small, fixed value ($10^{-9}$) while maintaining its direction. This allows us to measure the exponential divergence rate reliably. The leading Lyapunov exponent is calculated as}

\CRCAr{
\begin{equation}
    \lambda^{LE} = \frac{A}{t},
\end{equation}
where \( A \) is the cumulative sum of the logarithmic growth of the separation, and \( t \) is the total time. After each rescaling, \( A \)  is updated based on the ratio of the norms before and after rescaling. }
%Since the separation is rescaled every 10 time units, \( A \) is updated as:
% \begin{equation}
%     A = A + \ln \left( \frac{A_{dy}}{A_{dx}} \right)
% \end{equation}

% where \( A_{dx} \) is the norm before rescaling and \( A_{dy} \) is the norm after evolving for 10 time units. This method ensures accurate tracking of divergence despite resetting the perturbation magnitude.

\CRCAr{To ensure the calculation of the Lyapunov exponent is meaningful, we evolve four different initial conditions that lie on the attractor (see figure \ref{Lyapunov}c). The system is evolved for a sufficiently long time, and the LLE is updated incrementally until convergence with a tolerance of $10^{-6}$. The average of these independent calculations  yields  a leading Lyapunov exponent of $\lambda^{LE}=0.0329$. This value compares well with the leading Lyapunov exponent   predicted by DManD.}

% It is important to note that there is no available information on Lyapunov exponents for the DNS, so we cannot directly compare the DManD model predictions with DNS data.   

%-----------------------------------------
\subsection{Finding ECS in the model and DNS}
%-----------------------------------------

In this section, we leverage the DManD model with a dimension of $d_h = 20$ to explore the state-space of the low-dimensional representation and discover new ECS in the DNS. The primary goal is to use DManD to identify optimal initial conditions that can be fed into Openpipeflow,  which is equipped with an ECS solver for the full-state space. It is essential to highlight that discovering suitable initial conditions is pivotal in the success of
%to guarantee the efficacy of 
any ECS solver for high-dimensional systems (as we described in the introduction).

First, we summarise the approach that we use to find ECS within the context of DManD. This method follows the framework outlined by \citet{cvitanovic2005chaos}, and has been previously used by \citet{linot_Couette}. 
To identify ECS in the full state-space (which, despite being high-dimensional due to the numerical discretization of the infinite-dimensional NSE, is still finite), our objective is to find an initial condition that leads to a trajectory repeating over a defined time interval $T$. Thus, we aim  to search for solutions where the trajectory's behavior is periodic, essentially involving the identification of zeros in 
\begin{equation}
\boldsymbol{F} (\boldsymbol{v},T) =\boldsymbol{F}_T(\boldsymbol{v})-\boldsymbol{v},
\end{equation}
here, $\boldsymbol{F}_T(\boldsymbol{v})$ refers to the flow map $T$ time units from $\boldsymbol{v}$ (i.e., $\boldsymbol{F}_T(\boldsymbol{v}(t))=\boldsymbol{v}(t+T)$).
In manifold coordinates, this equation is expressed as
\begin{equation}
\boldsymbol{H} (\boldsymbol{h},T) =\boldsymbol{G}_T(\boldsymbol{h})-\boldsymbol{h},
\label{ECS_eq1}
\end{equation}
here, $\boldsymbol{G}_T(\boldsymbol{h})$ is the flow map $T$ time units from $\boldsymbol{h}$ (i.e., $\boldsymbol{G}_T(\boldsymbol{h}(t))=\boldsymbol{h}(t+T)$). To compute, $\boldsymbol{G}_T$, we use equation \ref{eq:Node2}.
Solving equation \ref{ECS_eq1} requires finding  both a  point $\boldsymbol{h}^*$ on the periodic orbit and the period $T^*$, such as $\boldsymbol{H} (\boldsymbol{h}^*,T^*) =0$. We use a Newton-Raphson method to determine $\boldsymbol{h}^*$ and $T^*$.

By taking  a Taylor series expansion  of $\boldsymbol{H}$, we find that near the fixed point $\boldsymbol{h}^*$, $T^*$ of $\boldsymbol{H}$, such that

\begin{equation*}
\left.\begin{matrix}
\boldsymbol{H} (\boldsymbol{h}^*,T^*)- \boldsymbol{H} (\boldsymbol{h},T) \approx \boldsymbol{D}_h\boldsymbol{H} (\boldsymbol{h},T)\Delta\boldsymbol{h}+\boldsymbol{D}_T\boldsymbol{H} (\boldsymbol{h},T)\Delta T,
\\ 
\boldsymbol{H}(\boldsymbol{h},T) \approx \boldsymbol{D}_h\boldsymbol{H} (\boldsymbol{h},T)\Delta\boldsymbol{h} + \boldsymbol{g}(\boldsymbol{G}_T(\boldsymbol{h}))\Delta T
\end{matrix}\right\}
\end{equation*}

\noindent
where $\boldsymbol{D}_h$ and $\boldsymbol{D}_T$ represent the Jacobians of $\boldsymbol{H}$ with respect to $\boldsymbol{h}$ and the period $T$, respectively. Additionally, $\Delta T = T^* - T$ and $\Delta \boldsymbol{h} = \boldsymbol{h}^* - \boldsymbol{h}$. We impose the constraint that the updates of $\Delta \boldsymbol{h}$ are orthogonal to the vector field $\boldsymbol{h}$ (i.e., $\boldsymbol{g}(\boldsymbol{h})^T \Delta \boldsymbol{h} = 0$). At a Newton step $(i)$, the system of equations becomes: 

\begin{equation}
\begin{bmatrix}
 \boldsymbol{D}_h\boldsymbol{H} (\boldsymbol{h}^{(i)},T^{(i)}) &  \boldsymbol{g}(\boldsymbol{G}_{T^{(i)}}(\boldsymbol{h}^{(i)})) \\ 
 \boldsymbol{g}(\boldsymbol{h}^{(i)})^T & 0
\end{bmatrix}
\begin{bmatrix}
\Delta\boldsymbol{h}^{(i)} \\ 
\Delta T^{(i)}
\end{bmatrix}
= -
\begin{bmatrix}
\boldsymbol{H} (\boldsymbol{h}^{(i)},T^{(i)})  \\ 
 0 
\end{bmatrix},
\end{equation}

\noindent

here, the standard Newton-Raphson method  updates the guesses $\boldsymbol{h}^{(i+1)}=\boldsymbol{h}^{(i)} + \Delta \boldsymbol{h}^{(i)}$ and $T^{(i+1)}=T^{(i)} + \Delta T^{(i)}$.

\begin{figure}
\begin{center}
\includegraphics[width=\linewidth]{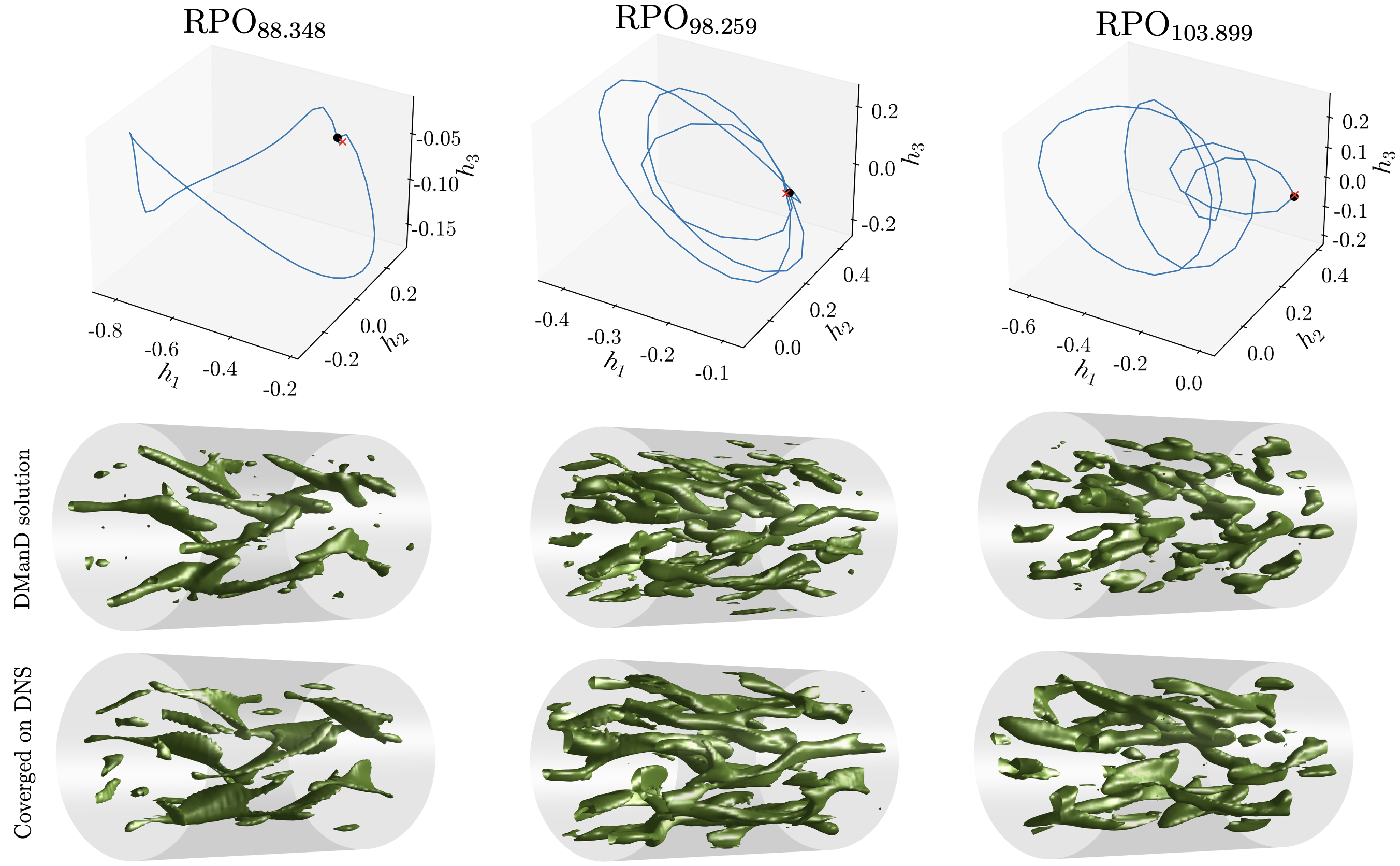}\\
% \begin{tabular}{cc}
% \includegraphics[width=0.53\linewidth]{Fig/DManD/PO2.png}&
% \includegraphics[width=0.48\linewidth]{Fig/DManD/EID_98.pdf}\\
% (a) & (b)\\
% \includegraphics[width=0.5\linewidth]{Fig/DManD/IC_DManD.png}&
% \includegraphics[width=0.47\linewidth]{Fig/DManD/IC_DNS.png}\\
% (c) & (d)\\
% \end{tabular}
\end{center}
\caption{\label{RPO_98d259_IC} 
The top panels show the converged ECS by DManD  
% (under a certain tolerance) 
projected onto the first three manifold coordinates.
The black dot and red diamond indicate the starting and ending points of the trajectory, respectively.
The middle and bottom panels show  snapshots of the vorticity flow field using the $\varlambda_2$ criterion with a threshold of $\varlambda_2=0.1$.
The middle panels represent  the ECS  from the DManD search, while the bottom panels display the converged state  from  the ECS solver of Openpipeflow.   Note that for representation purposes, we display $m_p=1$ instead of $m_p=4$.
% Panel (a) shows the RPO$_{98.259}$ discovered by DManD projected onto the first three manifold coordinates. The black dot and red diamond indicate the starting and ending points of the trajectory, respectively. Panel (b) display the power input versus dissipation of the RPO found in the DManD model at $d_h = 20$ and DNS.  Panels (c) and (d) show the converged state space of the RPO for the DManD and DNS.  Each snapshot displays the streamwise vorticity isosurfaces at $\omega_z=\pm 0.5913$ (red and blue, respectively). It is noted that for representation purposes, we display $m_p=1$ instead of $m_p=4$.
} 
\end{figure}

\begin{table}
    \centering
    \setlength{\tabcolsep}{8pt} % Adjust the column separation
    \begin{tabular}{ccccccc}
  Solution  & $\bar{D}$ & $\bar{c}$  & $\mu^{max}$ & \\%& Error \\ %& & Error-2  \\
  TW$_{1.332}$  & $1.3325$ & $0.93571$   & $ 0.04056$ &  \\%& $2.14\times10^{-6}$ \\ %& \\ %T5_IC211 0.749
  TW$_{1.798}$  & $1.7985$ & $1.6619$   & $ 0.08803$ &  \\%& $9.12\times10^{-4}$ \\ %& \\ %T5_IC67  0.602
  
  TW$_{1.448}$  & $1.4484$ & $1.8009$   & $ 0.04892$ &  \\%&  \\ %& & $7.28\times10^{-2}$\\ %T5_IC4
  TW$_{1.575}$  & $1.5759$ & $2.7052$   & $ 0.12259$ &  \\%&  \\ %& &$6.91\times10^{-2}$ \\ %T5_IC138 0.691
  RPO$_{4.891}$  & $1.4740$ & $2.7430$   & $ 0.06686$ &  \\%& \\ %& & $2.76\times10^{-3}$ \\ %T5_IC3
  RPO$_{19.631}$  & $2.0324$ & $3.30392$  & $ 0.03060$ & \\% & \\ %& & $2.34\times10^{-2}$\\ %T20_26
  RPO$_{19.631}$  & $2.0172$ & $3.30392$   & $ 0.03605$ & \\% & \\ %& & $2.38\times10^{-2}$ \\ %T20_157
  RPO$_{24.956}$  & $1.8644$ & $2.4739$   & $ 0.03784$ & \\% & \\ %& & $2.46\times10^{-2}$ \\ %T20_109
  RPO$_{35.042}$  & $1.7582$ & $1.6211$   & $ 0.04585$ & \\% & \\ %& & $1.30\times10^{-2}$ \\ %T40_IC335
  RPO$_{36.157}$  & $1.7503$ & $1.8426$   & $ 0.03465$ & \\% & \\ %& & $9.73\times10^{-2}$ \\ %T40_IC301
  RPO$_{38.654}$  & $2.0475$ & $1.142$   & $ 0.04499$ & \\% &  \\ %& &$6.11\times10^{-2}$ \\ %T40_IC267
  RPO$_{88.348}$  & $1.7985$ & $0.83400$  & $ 0.04039$ & \\% &  \\ %& & $1.15\times10^{-2}$ \\ %T90_IC30 0.01153
  RPO$_{94.891}$  & $1.9147$ & $4.01616$   & $ 0.03011$ &  \\%&  \\ %& $& $6.09\times10^{-2}$ \\ %T90_9 -0.06091
  RPO$_{98.259}$  & $2.0036$ & $1.5157$   & $ 0.02991$ & \\% &  \\ %& $& $3.98\times10^{-3}$ \\ %T100_3  0.003943
RPO$_{102.683}$  & $1.9793$ & $0.98244$   & $ 0.02988$ & \\% &  \\ %& & $1.98\times10^{-3}$ \\ %T100_v3_12 
  RPO$_{103.899}$  & $1.9036$ & $6.1943$  & $ 0.01088$ & \\% &  \\ %& & $2.15\times10^{-2}$ \\ %T100_33  -0.04258
  RPO$_{105.466}$  & $1.8270$ & $3.0258$  & $ 0.01653$ &  \\%&  \\% %& $4.25\times10^{-2}$ \\ %T100_v3_6
    \end{tabular}
    \caption{List of new invariant solutions for pipe flow at $Re=2500$ using initial conditions from the DManD model. The travelling waves are labelled with their dissipation rate $\bar{D}$, whereas the RPO are labelled by their period $T$. For each solution, we report  the average rate of dissipation $\bar{D}$, average downstream velocity $\bar{c}$,
    % the  dimension of the unstable manifold $d_U$,
    and the real part of the largest stability eigenvalue/Floquet exponent $\mu^{max}$. % Finally, the last two columns represent  `Error-1' for the Newton solver from the openpipeflow solver given as $\left \|  \boldsymbol{F}(\textbf{u}(T)) -\textbf{u}(0) \right \| / \left \| \textbf{u}(0) \right \|$;    % `Error-2' denotes the discrepancy in periods between the predicted RPOs from the DNS and DManD, given as  $\left \| T_{\text{DManD}} - T_{\text{DNS}} \right \| / \left \| T_{\text{DManD}} \right \|$ (similarly for TWs, we compare the $\bar{D}$).  
    }
    \label{ECS_tab}
\end{table}

\begin{figure}
\begin{center}
\begin{tabular}{c}
\includegraphics[width=\linewidth]{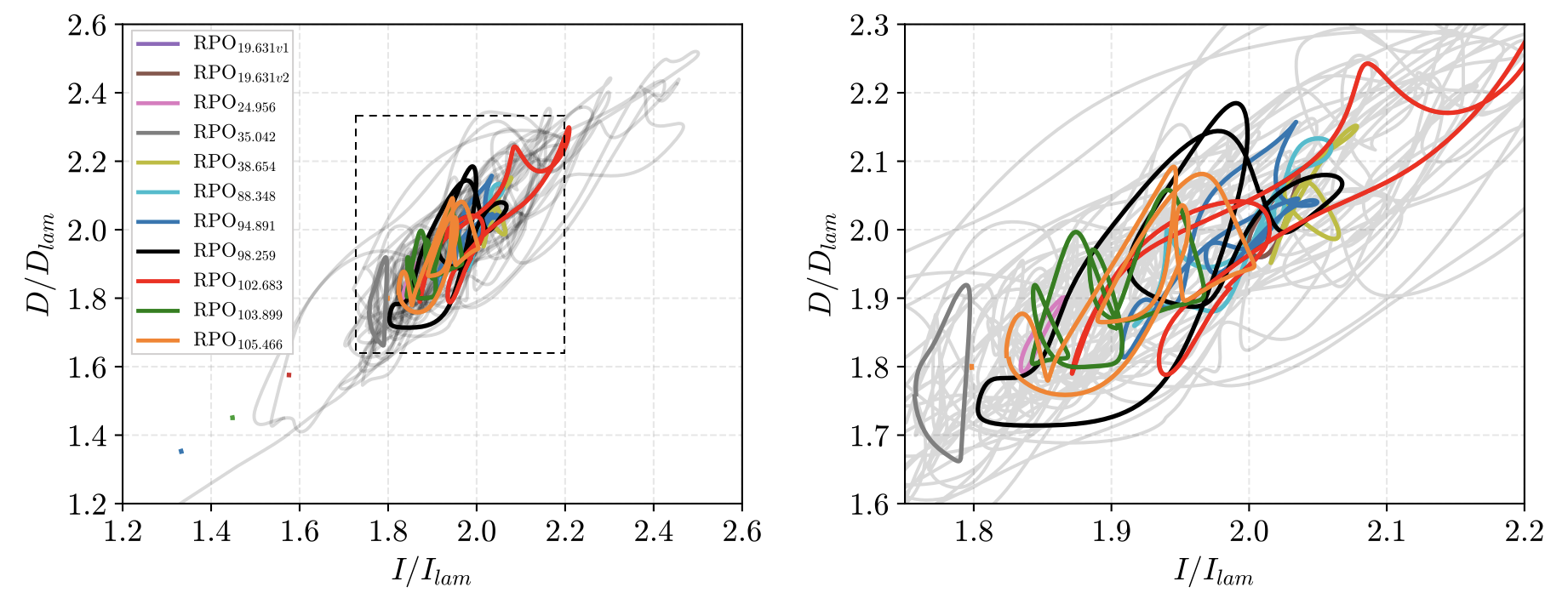}
\\
\end{tabular}
\end{center}
\caption{\label{ECS} 
The normalized dissipation verses power input for the collection of invariant solutions displayed in Table 2 with a long trajectory of the DNS turbulence plotted in the background. The dashed box in the left panel outlines the region that is magnified in the right panel for a closer view. } 
\end{figure}

A Newton scheme can be used to find ECS  within high-dimensional data,  the computational challenge posed by the Jacobian calculations has prompted the development of various solutions \citep{page_kerswell_2020,Page_2021,linot_Couette,Parker_2022,Yasuda}.  Openpipeflow addresses this issue by using a Jacobian-Free Newton-Krylov solver with a Hookstep-trust-region approach, as detailed by \citet{willis2019equilibria}. This solver efficiently bypasses the need for explicitly calculating the Jacobian when evaluating the objective function  $\boldsymbol{F}$. 
% \textcolor{black}{For the exploration of long orbits in pipe flow, \citet{budanur_jfm_2017} introduced a multiple-shooting method that is successful in identifying relative periodic orbits with periods up to $T \approx 68$.}  
\CRCAr{For a more detailed understanding of the ECS solver in the Openpipeflow solver, readers are referred to  \citet{willis2019equilibriaperiodicorbitscomputing}.}
The advantage of constructing a low-dimensional model that accurately captures dynamics becomes pivotal in the search for new ECS, as demonstrated in this section. Leveraging the inherent low dimensionality of DManD models, we choose to compute the Jacobian $\boldsymbol{D}_h\boldsymbol{H} (\boldsymbol{h},T)$ directly using the automatic differentiation tools employed during the training of the NODE. %To find these ECS, we use the SciPy `hybr' method, based on a modification of the Powell hybrid method by \citet{virtanen2020scipy}.

Our approach involves randomly selecting 100 initial conditions and exploring five distinct periods $T=[5,20,40,90,100]$ \CRCAr{(e.g., 500 guesses in total). In the reduced model, a Newton residual threshold of $10^{-3}$ identifies converged ECS candidates efficiently. These provide accurate initial guesses for the full system, where exact convergence is achieved with residuals between $10^{-4}$- $10^{-6}$, as used in previous works such as \citet{willis_avila_2013}.}

% ECS in the low-dimensional representation that have converged are then mapped back to the full space. The criterion for mapping the ECS from the low-dimensional model is a  Newton scheme residual of $10^{-2}$. While this residual may not be small enough for precise solutions, it provides a good initial guess for the DNS.  These initial conditions are subsequently used in Openpipeflow, employing its Newton solver for 100 iterations. It is important to note that the residual of the Newton scheme is reported, providing insights into the convergence behavior of the solver (we highlight that the success criterion for longer orbits is approximately $10^{-4}$, as used in previous works such as \citet{willis_avila_2013}).

As an illustrative example, the top panels of figure \ref{RPO_98d259_IC} depict the trajectories of three relative periodic orbits discovered by DManD, projected onto the three leading manifold coordinates $(h_1,h_2,h_3)$. %The starting and ending points of the trajectory are denoted by a dot and a diamond, respectively. We remark again that 
% These `discovered' ECS  are good initial conditions for the ECS solver in Openpipeflow. 
We also display the 
associated vorticity fields by displaying the $\varlambda_2$ criterion \citep{Jeong_Hussain_1995} for the state of the  converged ECS from DManD (middle panels) and the converged ECS in the DNS (bottom panels).
There is a notable qualitative agreement between these solutions, underscoring the effectiveness of using  converged ECS from DManD as robust initial conditions for the DNS ECS solver.

% Figure \ref{RPO_98d259_IC}b shows the ID plot for the new RPO discovered by DManD and the Openpipeflow solver. The predicted RPO follows a similar paths, but it cannot predict the high values of input and dissipation. However, to highlight that the converged solution in DManD provides with a good initial condition for the search for these solutions, we  present  three-dimensional snapshots of the state space from the relative periodic orbit attained in DManD, and the discovered on the DNS ECS solver (shown in figure \ref{RPO_98d259_IC}c and d, respectively).  We note the remarkable qualitative agreement between both solutions, which highlights  the success of utilizing converged solutions from DManD as robust initial conditions for the DNS ECS solver.

\begin{figure}
\begin{center}
\begin{tabular}{c}
\includegraphics[width=0.7\linewidth]{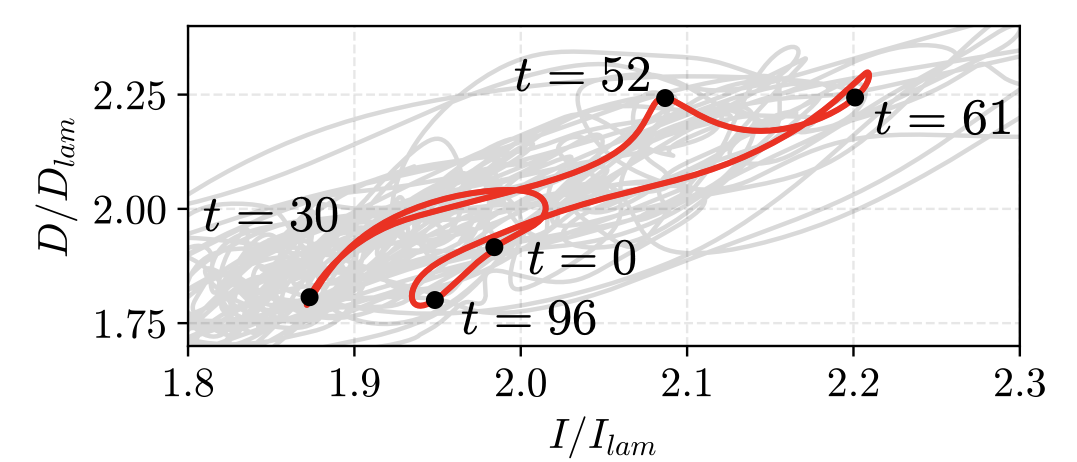}\\
\includegraphics[width=0.9\linewidth]{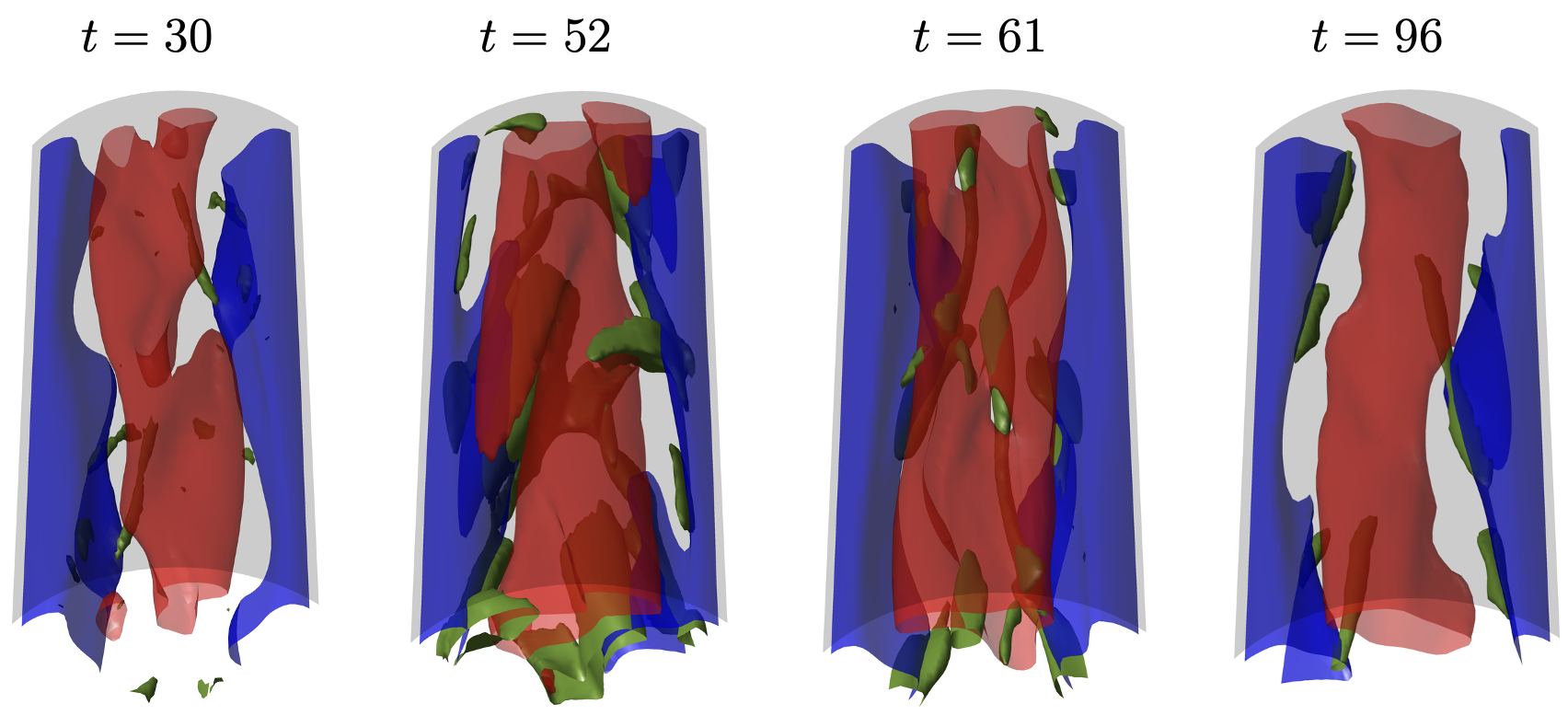}
\end{tabular}
% \begin{tabular}{cccc}
% $t=0$ & $t=17$ & $t=83$ & $t=94$  \\
% \includegraphics[width=0.23\linewidth]{Revision/Fig14_t30_halfdomain_combined.png}&
% \includegraphics[width=0.22\linewidth]{Revision/Fig14_t52_halfdomain_combined.png}&
% \includegraphics[width=0.22\linewidth]{Revision/Fig14_t61_halfdomain_combined.png}&
% \includegraphics[width=0.22\linewidth]{Revision/Fig14_t96_halfdomain_combined.png}\\
% \end{tabular}
\end{center}
\caption{\label{RPO_102d683} 
The normalized dissipation versus power input plot for RPO$_{102.683}$ is depicted, with a long trajectory of the DNS turbulence plotted in the background. The black dot points denote the times along the orbit at which 3D snapshots are shown below. 
Each snapshot displays \textcolor{black}{isosurfaces of the streamwise velocity fluctuations  for $ u_z' = 0.075 $ (blue, representing fast streaks) and $ u_z' = -0.075 $ (red, representing low-speed streaks). Additionally, each snapshot includes isosurfaces of the $\lambda_2$ criterion with a threshold of $ \lambda_2 = 0.1 $, shown in green, to highlight vortical structures. Note that for representation purposes, only a quarter of the domain is shown.} 
}
\end{figure}

% \begin{figure}
% \begin{center}
% \begin{tabular}{c}
% \includegraphics[width=\linewidth]{Fig/DManD/T_94d891.png}
% \end{tabular}
% \end{center}
% \caption{\label{RPO_98d259} 
% The normalized dissipation versus power input plot for RPO$_{94.891}$ is depicted, with a long trajectory of the DNS turbulence plotted in the background. The black dot points denote the times along the orbit at which 3D snapshots are shown below. Each snapshot displays the streamwise vorticity isosurfaces at $\omega_z=\pm 0.7$ (red and blue, respectively). It is noted that for representation purposes, we display $m_p=1$ instead of $m_p=4$.} 
% \end{figure}

\begin{figure}
\begin{center}
\begin{tabular}{c}
\includegraphics[width=0.7\linewidth]{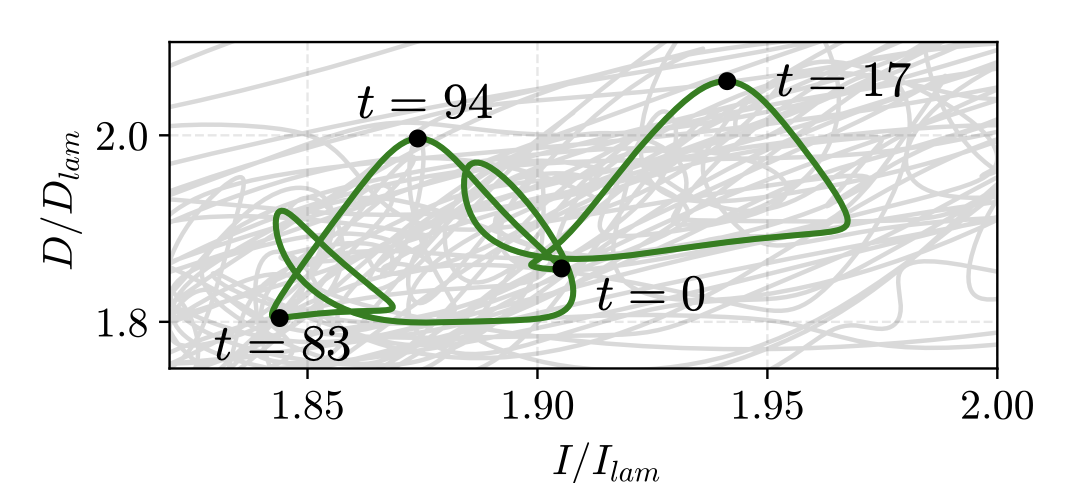}\\
\includegraphics[width=0.9\linewidth]{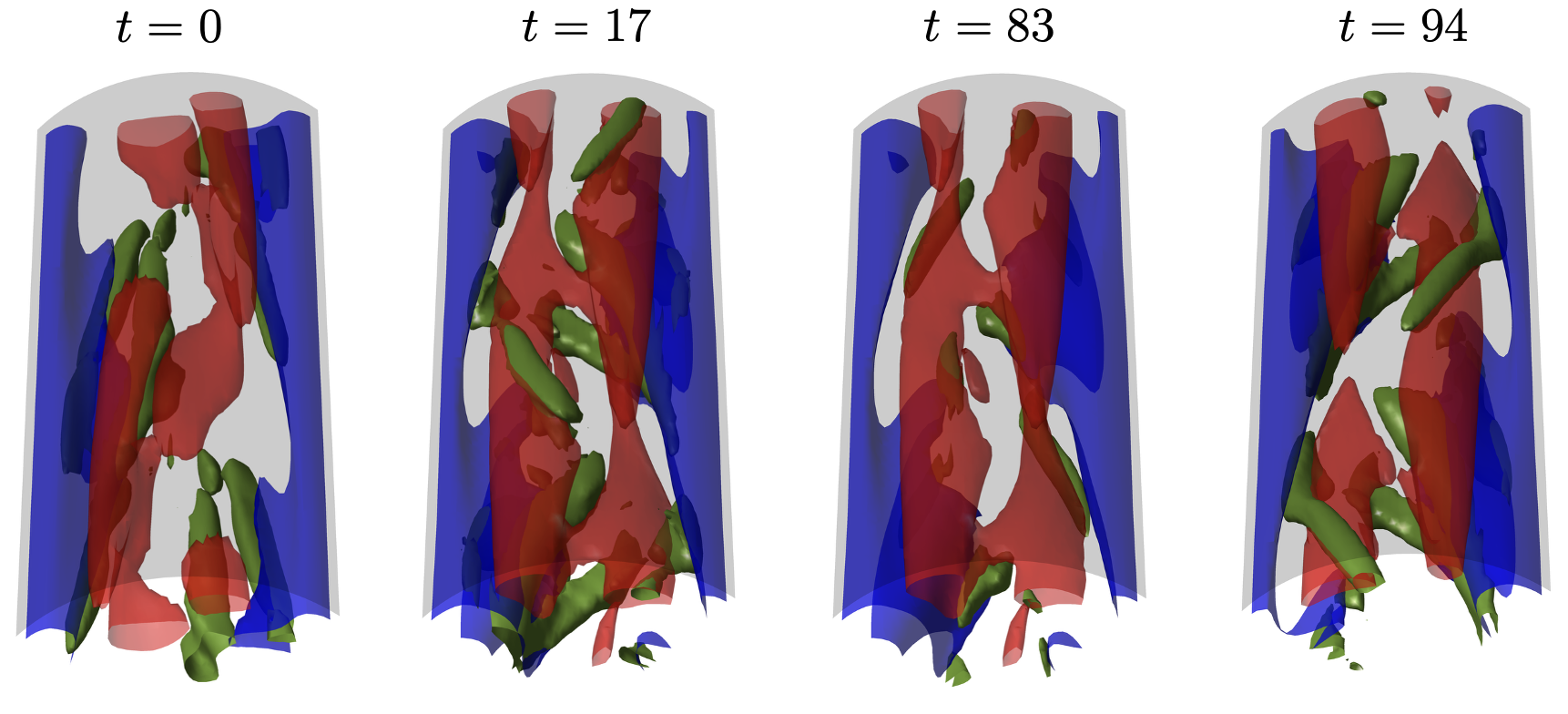}
\end{tabular}
% \begin{tabular}{cccc}
% $t=0$ & $t=17$ & $t=83$ & $t=94$  \\
% \includegraphics[width=0.22\linewidth]{Revision/t0_halfdomain_combined.png}&
% \includegraphics[width=0.22\linewidth]{Revision/t17_halfdomain_combined.png}&
% \includegraphics[width=0.22\linewidth]{Revision/t83_halfdomain_combined.png}&
% \includegraphics[width=0.22\linewidth]{Revision/t94_halfdomain_combined.png}\\
% \end{tabular}
\end{center}
\caption{\label{RPO_103d899} 
The normalized dissipation versus power input plot for RPO$_{103.899}$ is depicted, with a long trajectory of the DNS turbulence plotted in the background. The black dot points denote the times along the orbit at which 3D snapshots are shown below. 
Each snapshot displays \textcolor{black}{isosurfaces of the streamwise velocity fluctuations for $ u_z' = 0.075 $ (blue, representing fast streaks) and $ u_z' = -0.075 $ (red, representing low-speed streaks). Additionally, each snapshot includes isosurfaces of the $\lambda_2$ criterion with a threshold of $ \lambda_2 = 0.1 $, shown in green, to highlight vortical structures. Note that for representation purposes, only a quarter of the domain is shown.} 
% Each snapshot displays the $\varlambda_2$ criterion with a threshold $\varlambda_2=0.15$.Note that for representation purposes, we display $m_p=1$ instead of $m_p=4$.} 
}
\end{figure}

Table 2 presents 17 new ECS of pipe flow identified via the ECS solver embedded in Openpipeflow using the  converged solutions obtained via DManD. RPOs are labeled by their periods $T$ (in units of $\mathcal{D}/U$), while travelling waves (TWs) are labeled by their average dissipation rate ($\bar{D}$). Additionally, Table 2 includes information regarding the linear stability of each ECS, described by its Floquet multipliers $\Lambda = \exp(\mu_jT+i\theta_j)$. 
\CRCAr{
 We have performed a systematic cross-check of these ECSs  against all ECS from \citet{willis_avila_2013,Willis_2016} and \citet{budanur_jfm_2017}. We note that many of our newly discovered ECSs exhibit significantly longer periods than those previously reported (i.e., \citet{budanur_jfm_2017} reported RPOs with periods up to $T \approx 68$).}
% Some of these RPOs are characterized by the longest reported $T$ under these conditions, based on our current knowledge. 
% Table 2 also reports the absolute error between the ECS predicted by the DManD model and those found in the DNS by comparing their periods, and the dissipation rate for the RPO and TWs, respectively. 
% We observe a good agreement for the predicted ECS. 
We note that the DManD solver demonstrates its capability to accurately capture the orbit lengths of the ECS, with errors in their periods consistently below $1\%$. 
% In the remainder of the paper, we focus on representing the orbits from the ECS identified in the DNS solver.

One useful method for visualizing \CRCAr{of the high dimensional space of} these new invariant solutions is by projecting them onto the dissipation versus energy input plane. Figure \ref{ECS}  illustrates the  input and dissipation  projection of these new ECS, with
\CRCAr{the shaded region indicates the area in the input-dissipation projection where a long turbulent trajectory predominantly resides}.
% the darkest shading representing the DNS turbulent attractor.
It is expected that turbulence should explore more of the phase space and visit simple invariant solutions, then these ECS \CRCAr{appear to be embedded in this projection.
% should live in the core of the turbulent attractor.
While these solutions appear to reside within the central region of the $ID$ projection, determining their precise location within the turbulent attractor would require a more detailed analysis,   in the full state space, such as the work by \citet{Krygier}.  But we leave these avenue for future work.}
This finding is consistent with previous assertions suggesting that the turbulent attractor is guided by ECS \citet{chandler_kerswell_2013,Cvitanovic_2013,budanur_jfm_2017}.
The condition $D=I$ signifies dissipation balancing out energy input, which is the essential requirement for any equilibria or traveling wave (essentially an equilibrium in a co-moving frame of reference). We observe the discovery of four TWs.  The shaded region shows that DNS predominantly stays within the region of $1.6<D/D_{lam}<2.2$ and $1.7<I/I_{lam}<2.2$. Most of the discovered ECS remain within this region. In the right panel of figure \ref{ECS}, we present a magnified view \ref{ECS}. \CRCAr{This image reveals that many of these RPOs have complicated ID curves, whereas the periodic orbits exhibit simple loops.}

In figure \ref{RPO_102d683} and  \ref{RPO_103d899}, we focus on elucidating the state space of RPO$_{102.683}$ and RPO$_{103.899}$,
respectively.  The selection of
RPO$_{102.683}$ is based on its characteristic complex ID curve, which spans both high and low dissipation regions within the state space.
To understand the flow dynamics associated with this RPO, 
\textcolor{black}{ the bottom panels of
figure \ref{RPO_102d683} also shows  blue and red isosurfaces representing high ($u_z' = 0.075$)- and low ($u_z' = -0.075$)-speed streaks, respectively, while the green isosurfaces correspond to regions of high vorticity, identified using the $\lambda_2$ criterion with $\lambda_2=0.1$. 
We observe that the flow exhibits well-defined streamwise streaks during the entire cycle of the RPO, with some wavy disturbances at $t=30$ and $t=52$ and $t=96$. } At the points of higher dissipation $(t=52)$ and $(t=61)$, the vorticity appears to undergo significant shearing, leading to vortex breakup (see the intense isosurfaces associated with $\lambda_2$ criterion).
High rates of dissipation correspond to intense velocity gradients, and subsequently, it leads to 
 high values of the strain rate tensor. This indicates that these times are characterized by strong and turbulent flow structures.
At the lowest point of dissipation, see $t=30$ and $t=96$,  the vortex structures are much weaker and less pronounced. This weakening of the vortex structures correlates with lower dissipation rates.  Then, the dynamics stabilise due to the increases in the power input ($t=96$),
in comparison to $t=52$.
It is worth noting that for these 3D snapshots visualizing just one-quarter of the pipe. Finally, we turn attention to the RPO$_{103.899}$, the top panel of figure  \ref{RPO_103d899} presents the ID projection for this RPO, which intriguingly resembles the shadow of two shorter orbits, similar to the described by  \citet{budanur_jfm_2017} \CRCAr{(see their figure 11)}.
\textcolor{black}{We also observe well formed fast streaks in the domain.}
This results in the dynamics at the highest point of dissipation at  $t=17$ and $t=94$ the dynamics exhibit significant shearing, leading to intense vortex interactions and high dissipation rates. These moments are characterized by extreme shearing,  events \textcolor{black} {affecting the slow streaks}. Conversely, at the lowest point of dissipation, for example $t=83$, the vortex structures are notably weaker. This reduced dissipation corresponds to a more stable and less turbulent flow state.

\section{Conclusions \label{conclusion}}

In this study, we have built data-driven models for pressure-driven fluid flow through a circular pipe. To reduce the computational requirements, we impose the shift-and-reflect symmetry to study the system in a minimal computational cell at $Re=2500$. Nonetheless, this computational size is capable of maintaining the chaotic \CRCAr{nonlinear} dynamics of turbulence. 
To build these data-driven models, \CRCAr{we} employed DManD (Data-driven Manifold Dynamics), an invariant-manifold-based method. DManD is based on the idea of modelling of turbulence from a dynamical systems approach in which the long-time dynamics of the dissipative NSE are expected to live in a finite-dimensional invariant manifold. Thus, DManD allows the \CRCAr{parameterization} of the invariant manifold with vastly fewer degrees of freedom compared to the original data. For learning these manifold coordinates, we first perform a linear dimension reduction with POD, and then a nonlinear dimension reduction via  autoencoders which are capable of accurately \CRCAr{predicting}  the low POD coefficients.  Finally, we use a state-space approach with Neural ODEs (NODEs) within these learned coordinates to model the dynamics.  This combination of linear and nonlinear techniques allows for a compact and efficient representation of the turbulent flow dynamics.
Our framework, solely driven by data, enables us to construct models with fewer than 20 degrees of freedom, a significant reduction compared to fully resolved DNS that requires on the order of $\mathcal{O}(10^5)$. 
% We note that our claim is not related to the dimensional of the manifold, but to the fact that only 20 degrees of freedom are needed to effectively capture crucial features of the flow, including streak breakdown and the regeneration cycle.
In short-time tracking, they accurately track the true trajectory for one Lyapunov time.
\CRCAr{Additionally, the leading Lyapunov exponent estimated from DManD closely matches that obtained from the DNS, confirming that our approach captures the  chaotic dynamics and the short-term predictability of the flow.}
In the long term, the models successfully capture key aspects of the nonlinear dynamics such as Reynolds stresses and probability distribution in $ID$ space.

We have also identified seventeen previously unknown ECS for turbulent pipe flow at $Re=2500$. The success in discovering these new ECS lies in using converged ECS from DManD at $d_h=20$ as effective initial conditions for the ECS solver in Openpipeflow.   This approach has led to the reporting of RPOs with the longest periods observed for three-dimensional turbulent pipe flow to date, to the best of our knowledge. These periodic orbits are situated within the core of the state space traversed by the turbulent attractor. This finding is consistent with previous assertions suggesting that the turbulent attractor is guided by ECS \CRCAr{\citep{Hopf_1948,Cvitanovic_2013,budanur_jfm_2017,Page_2024}}.

% Overall, our findings demonstrate the efficacy of employing neural networks and state-space approaches in understanding and predicting turbulent flow dynamics, opening avenues for further exploration and advancements in the field of fluid mechanics.

Accurately modeling turbulent dynamics with significantly fewer degrees of freedom than required for DNS, as demonstrated by the manifold dynamics models presented here, opens up exciting possibilities for dynamical-systems type analyses. These models enable the calculation of local Lyapunov exponents in a computationally efficient manner. We also highlight that we have presented a global description of the manifold, but it would be possible to divide the global manifold topology into many local representations called charts \citep{Floryan,Fox}. 
\CRCAr{In experimental settings, temporal data is typically limited to partial observables (i.e., measurements at a few spatial locations) making full state methods such as DManD  difficult to implement. Nonetheless, data driven models constructed from such partial measurements may still enable control-oriented strategies. Investigating this direction represents a promising avenue for future research.}
% Finally, data-driven manifold-invariant models facilitate the application of control strategies with minimal degrees of freedom, \CRCAr{ particularly when derived from numerical datasets}, as exemplified by the use of reinforcement learning techniques in controlling plane MFU Couette flow \citet{Couette_control}.

Importantly, the models developed in this study are dependent on specific system parameters such as the Reynolds number. Therefore, when transitioning to different Reynolds numbers, it is necessary to obtain a new dataset to adjust the weights of the neural networks for the autoencoders and NODE. Therefore, a crucial direction for future research is to develop models that capture this parameter dependence.  The goal is to create robust low-dimensional models capable of transferring knowledge across different parameter regimes. 
Achieving this would enable broader applications and provide deeper insights into turbulent flow dynamics. This represents a key direction for future research in this field.\\

Declaration of Interests. The authors report no conflict of interest. \\

\textcolor{black}{ Data availability. The code and data (in compressed format) used in the paper is available in the group GitHub repository.
%https://github.com/haller-group/SSMLearnPy
The complete dataset is available from the authors upon request.}\\

We wish to thank A.P. Willis for making the ‘Openpipeflow’ code open source, and his help setting up the simulations. We gratefully acknowledge Jake Buzhardt for helpful discussions. This work was supported by  ONR N00014-18-1-2865 (Vannevar Bush Faculty Fellowship). \textcolor{black}{The authors gratefully acknowledge the valuable feedback provided by the three anonymous reviewers during the revision process.} \\

\bibliographystyle{jfm}
\bibliography{jfm-instructions.bib}

\end{document}